\documentclass{rstransa} 

\usepackage{microtype}

\usepackage[normalem]{ulem}

\usepackage{mathtools} 

\begin{document}

\title{Proper Orthogonal and Dynamic Mode Decomposition of Sunspot Data}

\author{ A. B. Albidah$^{1,2}$, W. Brevis$^{3}$, V. Fedun$^{4}$, I. Ballai$^{1}$, D. B. Jess$^{5,6}$, M. Stangalini$^{7}$, J. Higham $^{8}$ and G. Verth$^{1}$}


\address{$^{1}$Plasma Dynamics Group, School of Mathematics and Statistics, The University of Sheffield, Hicks Building, Hounsfield Road, Sheffield, S3 7RH, United Kingdom\\
$^{2}$ Department of Mathematics, College of Science Al-Zulfi, Majmaah University, Saudi Arabia\\ 
$^{3}$School of Engineering, Pontificia Universidad Cat\'{o}lica de Chile, Chile\\
$^{4}$Plasma Dynamics Group, Department of Automatic Control and Systems Engineering, The University of Sheffield, Sheffield, S1 3JD, United Kingdom\\
$^{5}$ Astrophysics Research Centre, School of Mathematics and Physics, Queen’s University, Belfast, BT7 1NN, UK \\
$^{6}$ Department of Physics and Astronomy, California State University Northridge, Northridge, CA 91330, USA \\
$^{7}$ ASI Italian Space Agency, Via del Politecnico snc, 00133, Rome, Italy \\
$^{8}$ School of Environmental Sciences, Department of Geography and Planning, University of Liverpool, Roxby Building, Liverpool, L69 7ZT, UK
}

\subject{Solar Physics, Mode Decomposition methods, MHD waves}


\corres{Gary Verth\\
\email{g.verth@sheffield.ac.uk}}

\begin{abstract}
 High resolution solar observations show the complex structure of the magnetohydrodynamic (MHD) wave motion. We apply the techniques of POD and DMD  to identify the dominant MHD wave modes in a sunspot using the intensity time series. The POD technique was used to find modes that are spatially orthogonal, whereas the DMD technique identifies temporal orthogonality. Here we show that the combined POD and DMD approaches can successfully identify both sausage and kink modes in a sunspot umbra with an approximately circular cross-sectional shape.
\end{abstract}




\maketitle
\section{Introduction}


 Analysis of oscillations in sunspot data began in the late 1960's, see e.g., Beckers and Tallant \cite{beckers1969chromospheric}. These authors determined the observational parameters of umbral flashes, a phenomenon that shows oscillatory behaviour in a sunspot. In the early 1970's several studies looked at Doppler velocity oscillations in sunspots. Bhatnagar \cite{bhatnagar1971oscillatory}  determined Doppler velocity oscillations with a period of the order of $180-220 s$. Later Beckers and Schultz \cite{beckers1972oscillatory} observed a peak period of around $180s$. Measuring intensity oscillations directly from time lapse filtergram movies, Bhatnagar and Tanaka \cite{bhatnagar1972intensity} detected periodicities of the order of $170 \pm 40s$. Later on, Moore \cite{moore1981dynamic} detected Doppler velocity oscillations of $120-180 s$ and $240-300 s$ in umbral and penumbral regions, respectively
 
 To the present day the study of oscillations in sunspots has mainly been carried out by Fourier transforming data to provide, e.g. power spectra, either on a pixel by pixel basis or integrating across a particular Region of Interest (ROI). Although such analysis can provide valuable information, for the identification of coherent structures, e.g.  magnetohydrodynamic (MHD) wave modes, in the temporal and spatial domain across a particular ROI, the basic Fourier transform approach has its limitations. Despite this, one can fine tune a Fourier filter in the spatial and temporal domains to try and identify particular MHD wave modes, as was presented by Jess et al. \cite{jess2017inside} (hereafter J17) in order to detect a slow kink body mode in a sunspot umbra. In the present work we aim to apply the more advanced techniques of Proper Orthogonal Decomposition (POD) and Dynamic Mode Decomposition (DMD) to identify low order MHD wave modes as coherent oscillations across the sunspot umbra, both in the spatial and temporal domains, using the same sunspot data as \cite{jess2013influence, jess2016solar, jess2017inside}. In the more general solar context, POD has previously been applied to decompose the Doppler velocity of the entire solar disc \cite{vecchio2005proper, vecchio2006full, vecchio2008spatio} and numerical convection data \cite{onofri2012propagation}.

\section{Observations}
\label{sec:obs}
\begin{figure}[htp]
  \centering
  \begin{tabular}{ c c c }
    \includegraphics[width=41mm]{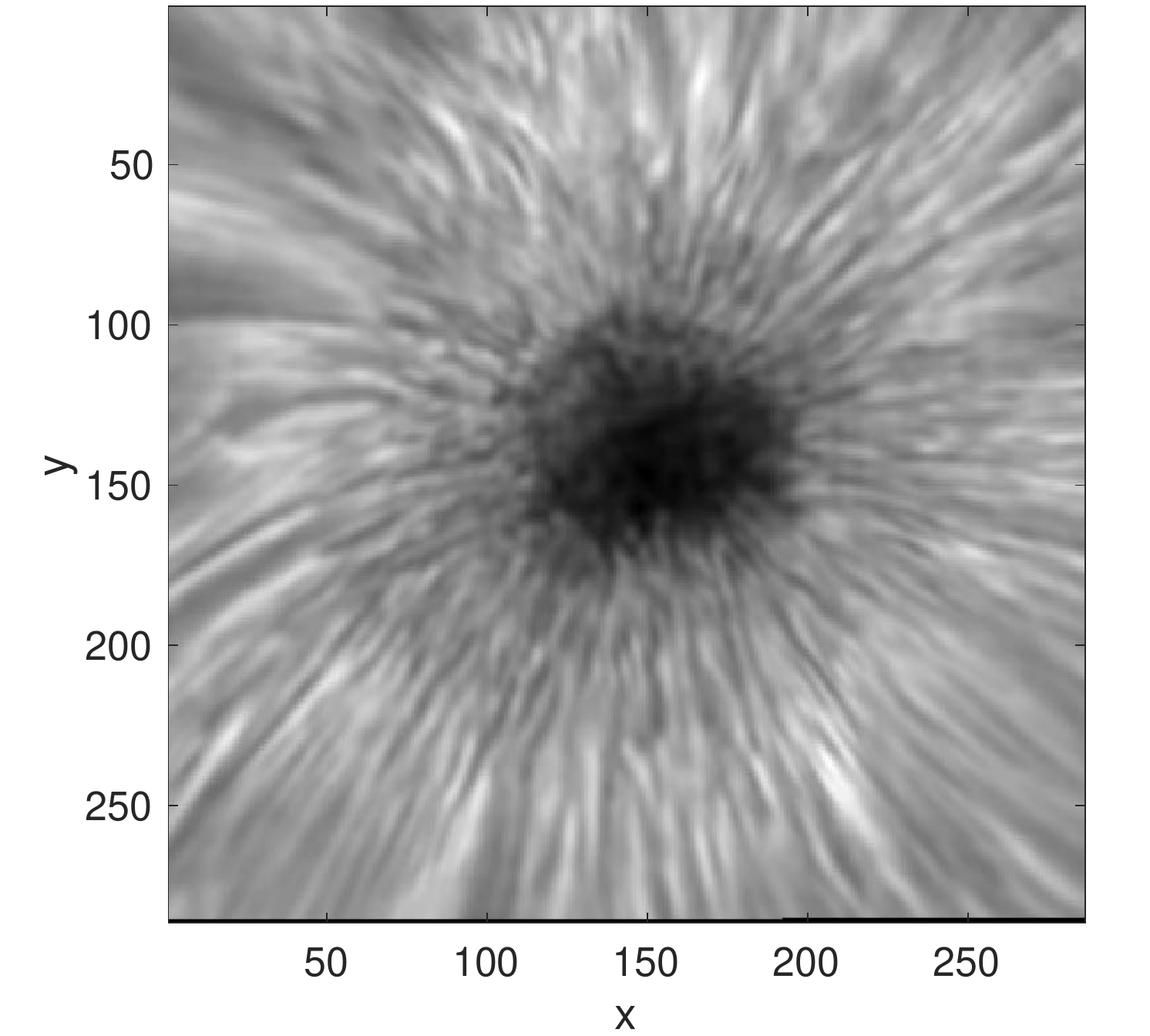}

    \includegraphics[width=43mm]{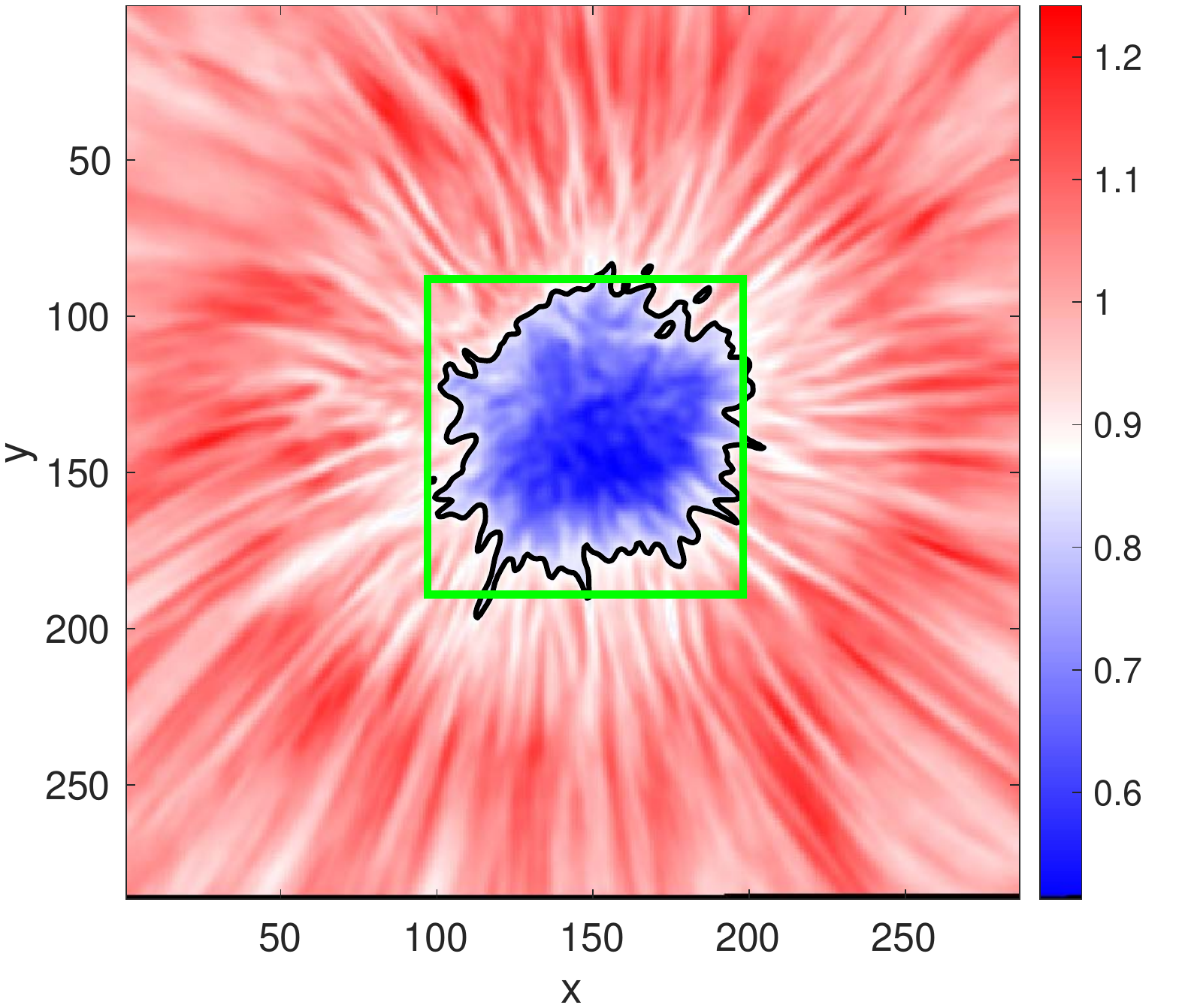}

    \includegraphics[width=46mm]{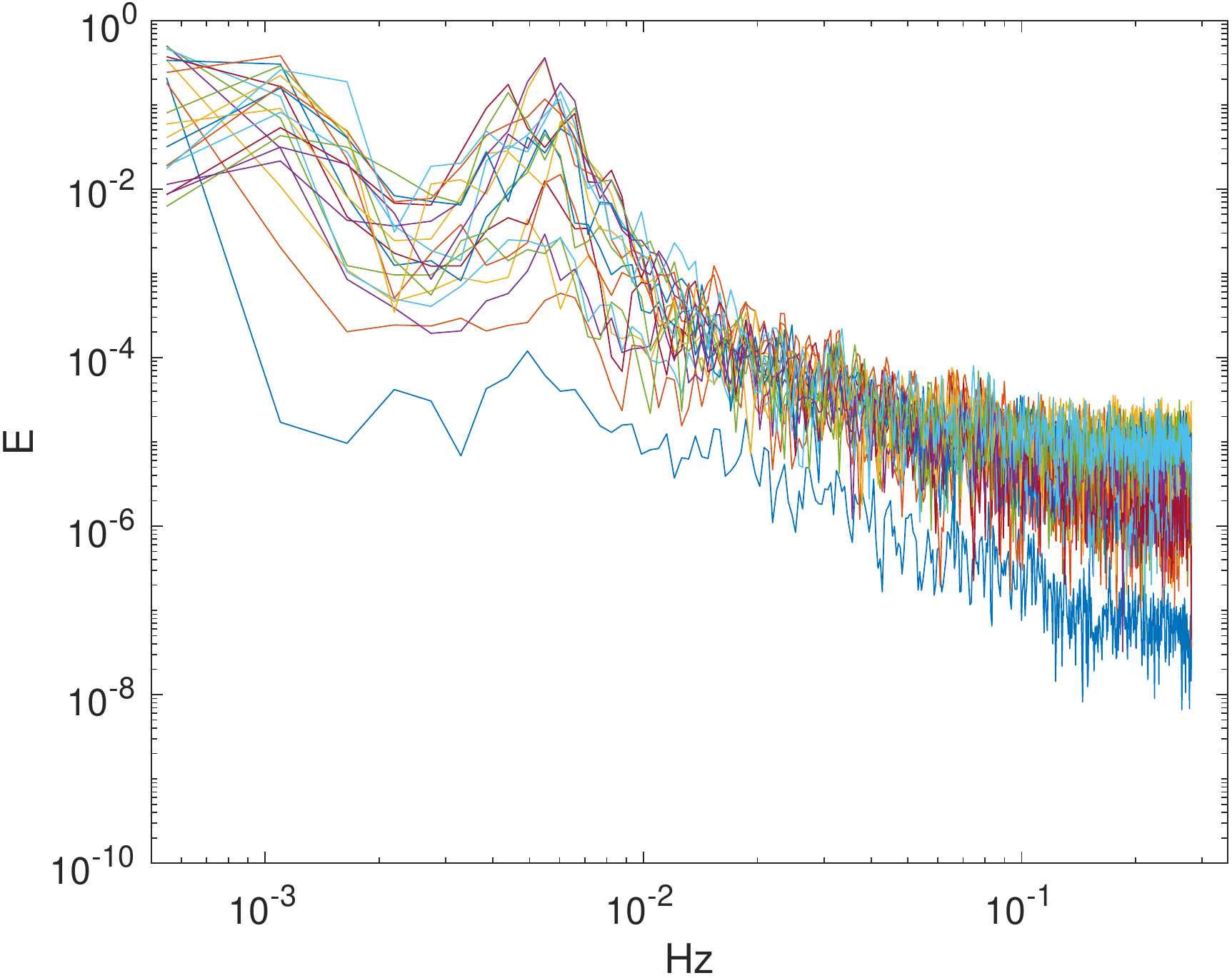}

  \end{tabular}
  \caption{The left panel shows a snapshot from the H$\alpha$ time series with the spatial scale in pixels (one pixel has a width of $0.138{^{\prime\prime}}$ which is approximately 100 km on the surface of the Sun). The middle panel shows the mean intensity of the time series, the colourbar displays the magnitude of the mean time series, the solid black line shows umbra/penumbra boundary (intensity threshold level 0.85) and the green box ($101 \times 101$ pixels) shows the region where we apply our POD and DMD analysis. The right panel displays the PSD of the time coefficients of the first 20 POD modes (in log scale). The PSD shows peaks between frequencies $4.3$ mHz and $6.5$ mHz (corresponding to periods of 153 - 232s). \label{fig:POD}}
 \end{figure}

The dataset we will analyse has been previously used for studies of running penumbral waves \cite{jess2013influence}, connections between photospheric and coronal magnetic fields \cite{jess2016solar} and in the detection of an umbral kink mode \cite{jess2017inside}. The portion of the complete multi-wavelength dataset used in the present study consists of a 75-minute observing sequence of H$\alpha$ images acquired by the Hydrogen-Alpha Rapid Dynamics camera (HARDcam;\cite{jess2012source}). The H$\alpha$ time series, which observed the approximately circular sunspot present within the active region NOAA~11366, were obtained during excellent seeing conditions between 16:10 -- 17:25~UT on 10 December 2011, with the Dunn Solar Telescope (DST) at Sacramento Peak, New Mexico. The sunspot under investigation was located at N17.9W22.5 in the conventional heliographic co-ordinate system, or $(356{^{\prime\prime}}, 305{^{\prime\prime}})$ in heliocentric co-ordinates. The filter employed had a full-width at half-maximum of 0.25{\,}{\AA}, which was centered on the H$\alpha$ line core at 6562.8{\,}{\AA}. A platescale of $0.138{^{\prime\prime}}$ per pixel was used to provide a field-of-view size equal to $71{^{\prime\prime}}\times71{^{\prime\prime}}$. On-site high-order adaptive optics \cite{rimmele2004first}, post-facto (speckle) image reconstruction techniques \cite{woger2008speckle} and image destretching relative to simultaneous broadband continuum images \cite{jess2010study} were implemented to improve the final data products, providing a cadence of 1.78{\,}s. A sample H$\alpha$ image of the sunspot is displayed in the left panel of Figure \ref{fig:POD}.

\section{Modal decomposition techniques} \label{sec:style}
Following the approach by Higham et al. \cite{higham2018implications}, we are going to employ two techniques to identify low order MHD modes from the intensity time series. The POD method can be used to identify MHD wave modes by imposing the criteria that modes are spatially orthogonal. The second method, DMD, assumes a temporal orthogonality of modes. Hence, if observed MHD wave modes do not have identical frequencies, and this difference is resolved in the frequency domain, then DMD offers an optimal methodology to identify such modes. 

\subsection{Proper Orthogonal Decomposition (POD)} \label{sec:POD}
The POD technique was developed by Pearson\cite{pearson1901liii} as an analogue of the principal axis theorem in mechanics. POD was introduced as a mathematical technique in fluid dynamics by Lumley \cite{lumley1967structure} to identify coherent structures in turbulent flow-fields. In the literature POD takes a variety of names depending on the field of application, such as principal component analysis (PCA) and Hotelling analysis. Since POD will produce as many modes as there are time snaphots in a dataset, the challenging part of this type of analysis is to identify which of the POD modes actually have a physical meaning. Hence, for identification of MHD wave modes in the umbral regions of sunspots care should be taken to compare POD modes with what we should expect from theoretical models, e.g. the MHD wave modes of cylindrical magnetic flux tubes.

Let us consider the sequence of ROI intensity snapshots of the sunspot of spatial size $X \times Y$ and a time domain of size $T$. Each of these snapshots can be reorganized in a column matrix $\boldmath{W} \in \mathbb{R}^{N \times T}$, where $N=XY$ and $N\gg T$, where each column of $\boldmath{W}$ will be defined as $\boldmath{w_{i}}$ with $i=1...T$ such that
\begin{equation}
    W=\{w_{1},w_{2},...w_{T}
    \}.
\end{equation}

There are several approaches to applying POD to a dataset. The classical POD method \cite{eckart1936approximation} is performed by computing the eigenvalues  and the eigenvectors of the covariance matrix of the dataset. 

Another approach is to obtain the POD of $\boldmath{W}$ using the optimum low rank approximation \cite{eckart1936approximation} and this is known as the Singular Value Decomposition (SVD). Applying the SVD, we obtain
\begin{equation}
    W=\Phi S C^{*}.
\end{equation}
This decomposition gives the spatial structure of each mode in the columns of the matrix $\boldmath{\Phi} \in \mathbb{R}^{N\times T}$, i.e. $\boldmath{\phi_{i}}$ with $i=1...T$ and these modes are orthogonal to each other. The temporal evolution of the POD modes are given by the columns of the matrix $\boldmath{C}\in \mathbb{R}^{T \times T}$. The particular spatial and temporal output of the POD presented here is the product of the $N$ two dimensional spatially orthogonal eigenfunctions with their associated  one dimensional time coefficients. Since POD places no restriction on the time coefficients, these can be periodic or aperiodic and the amplitude can also vary with time. The matrix $\boldmath{S}\in \mathbb{R}^{T \times T}$ is a diagonal matrix, and the modes are generally ranked according to their contribution to the total variance of the snapshot series. This contribution is given by the diagonal elements of matrix, $\boldmath{\lambda}$, by means of the vector $\boldmath{\lambda}= \mathrm{diag}(\boldmath{S})^{2}/(N-1)$. 

 \begin{figure}[htp]
  \centering
  \begin{tabular}{ccc}
    
    \includegraphics[width=40mm]{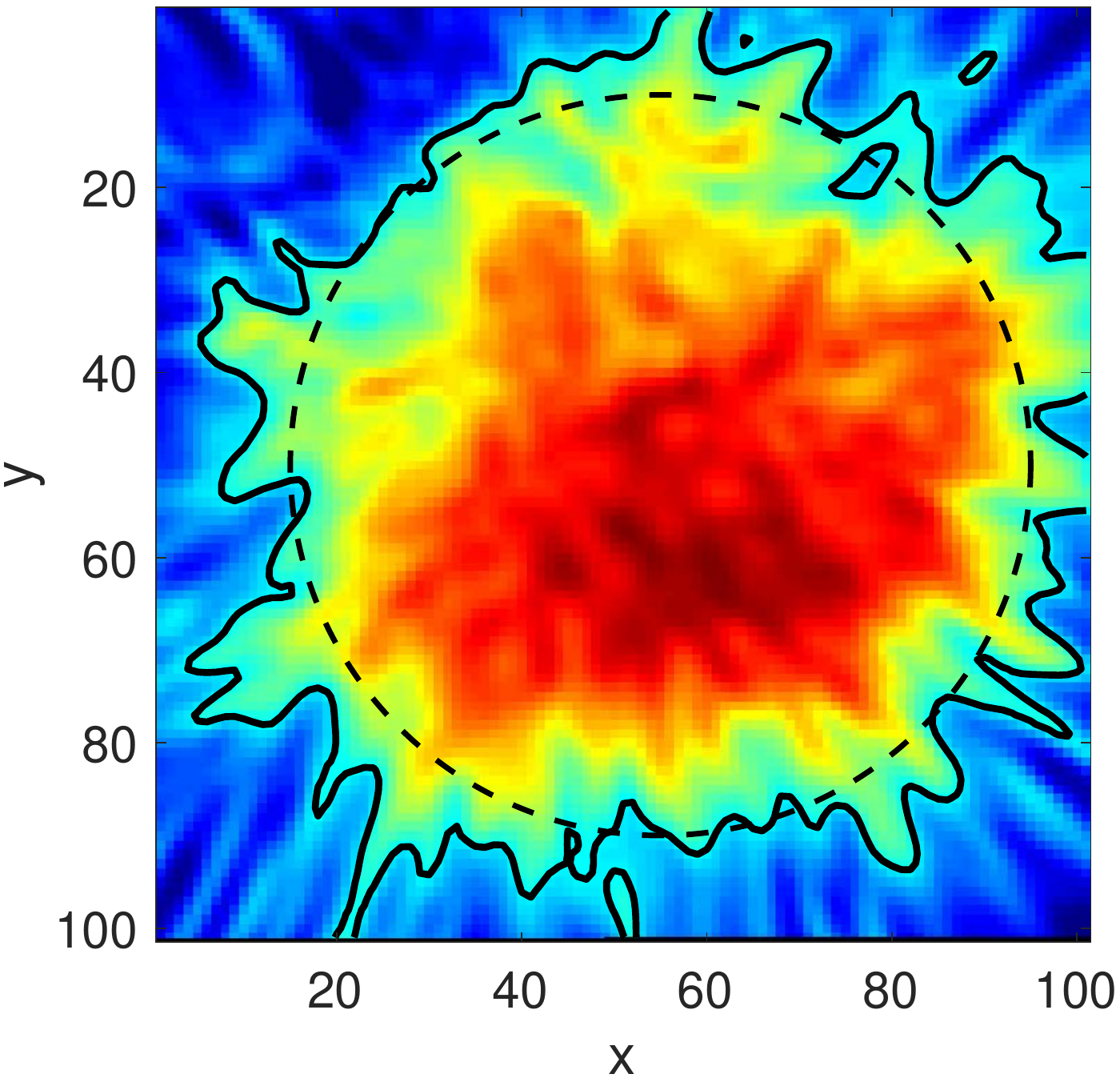}&

     \includegraphics[width=40mm]{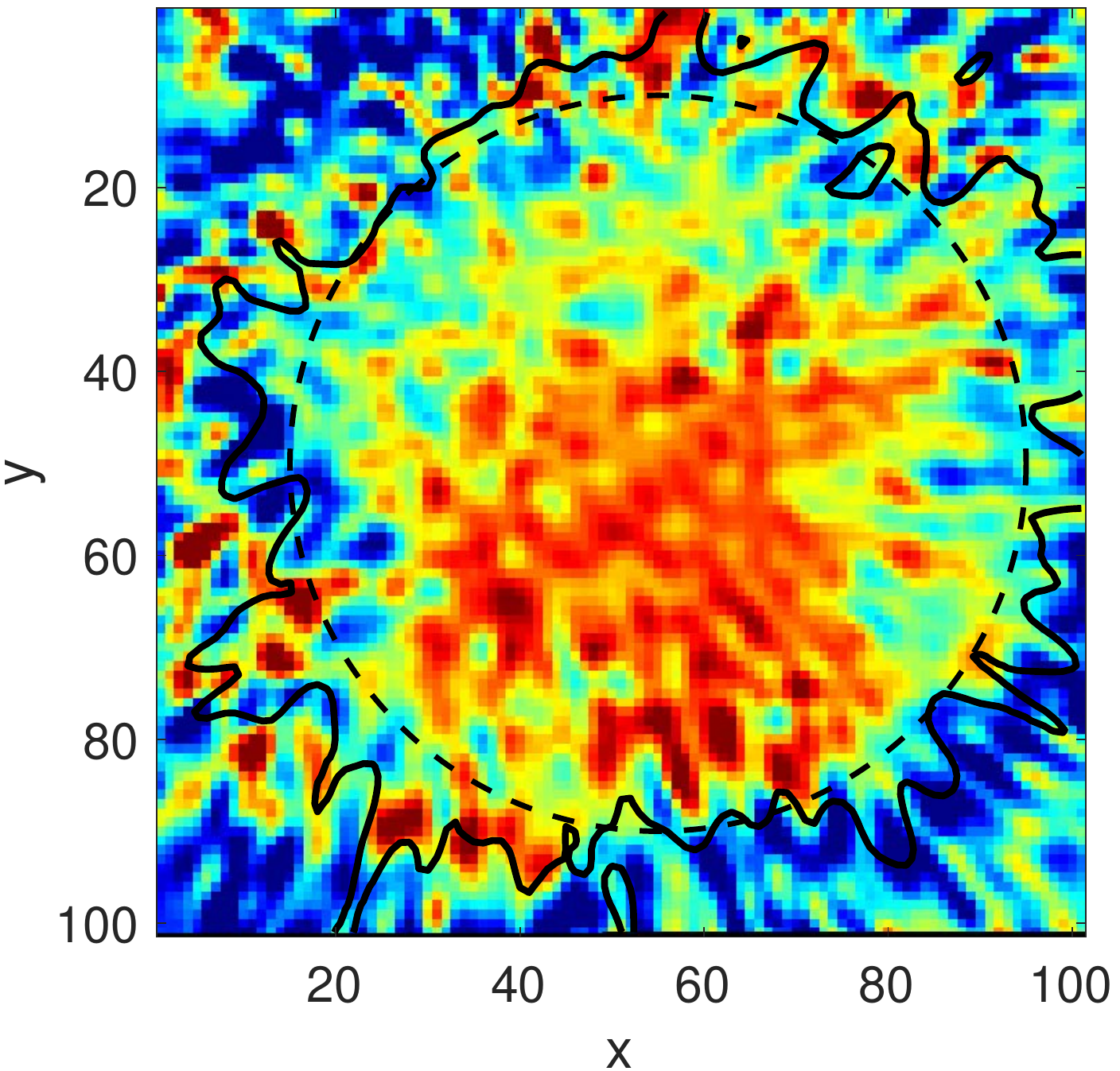}&

      \includegraphics[width=45mm]{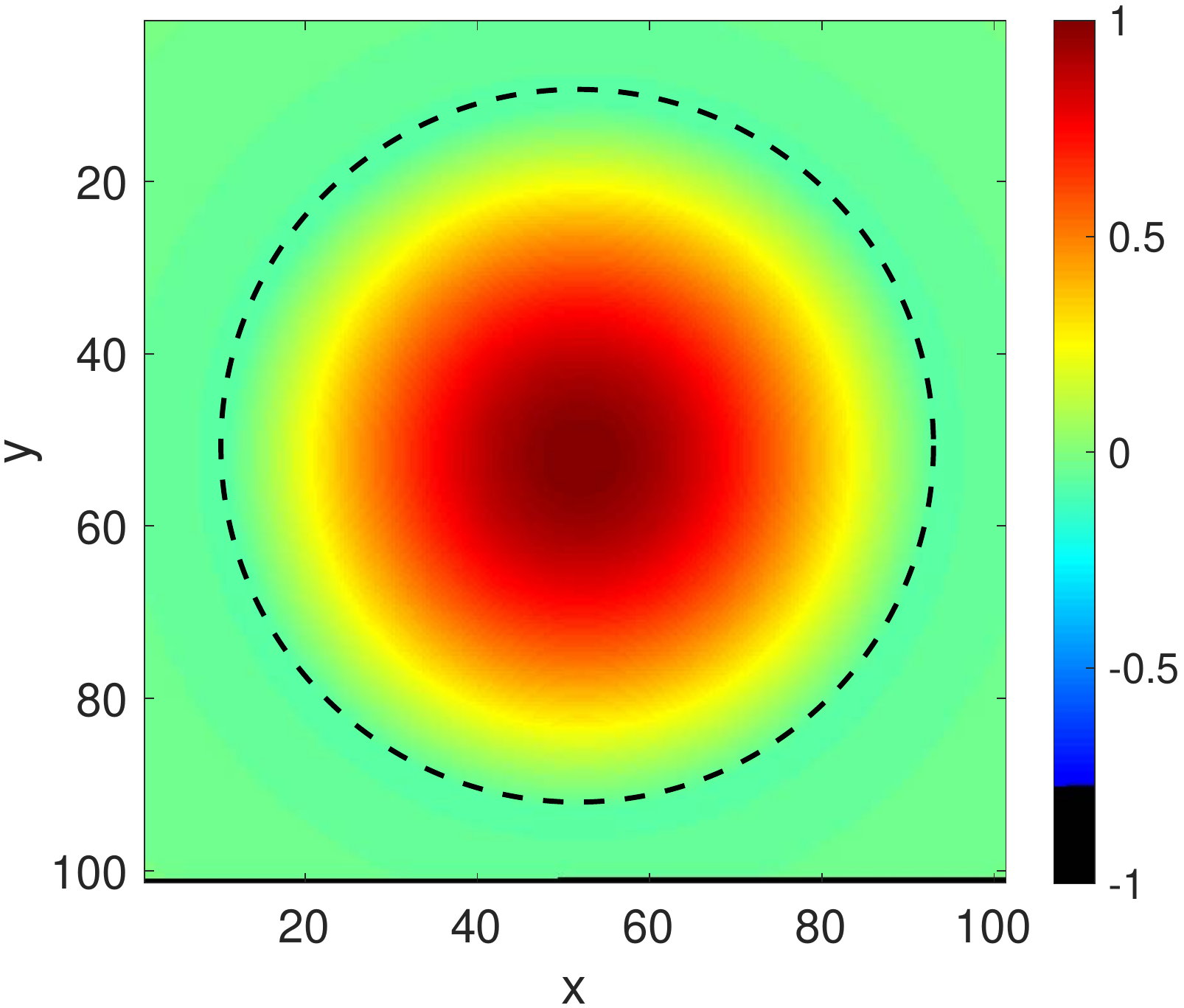}\\
     
    \includegraphics[width=40mm]{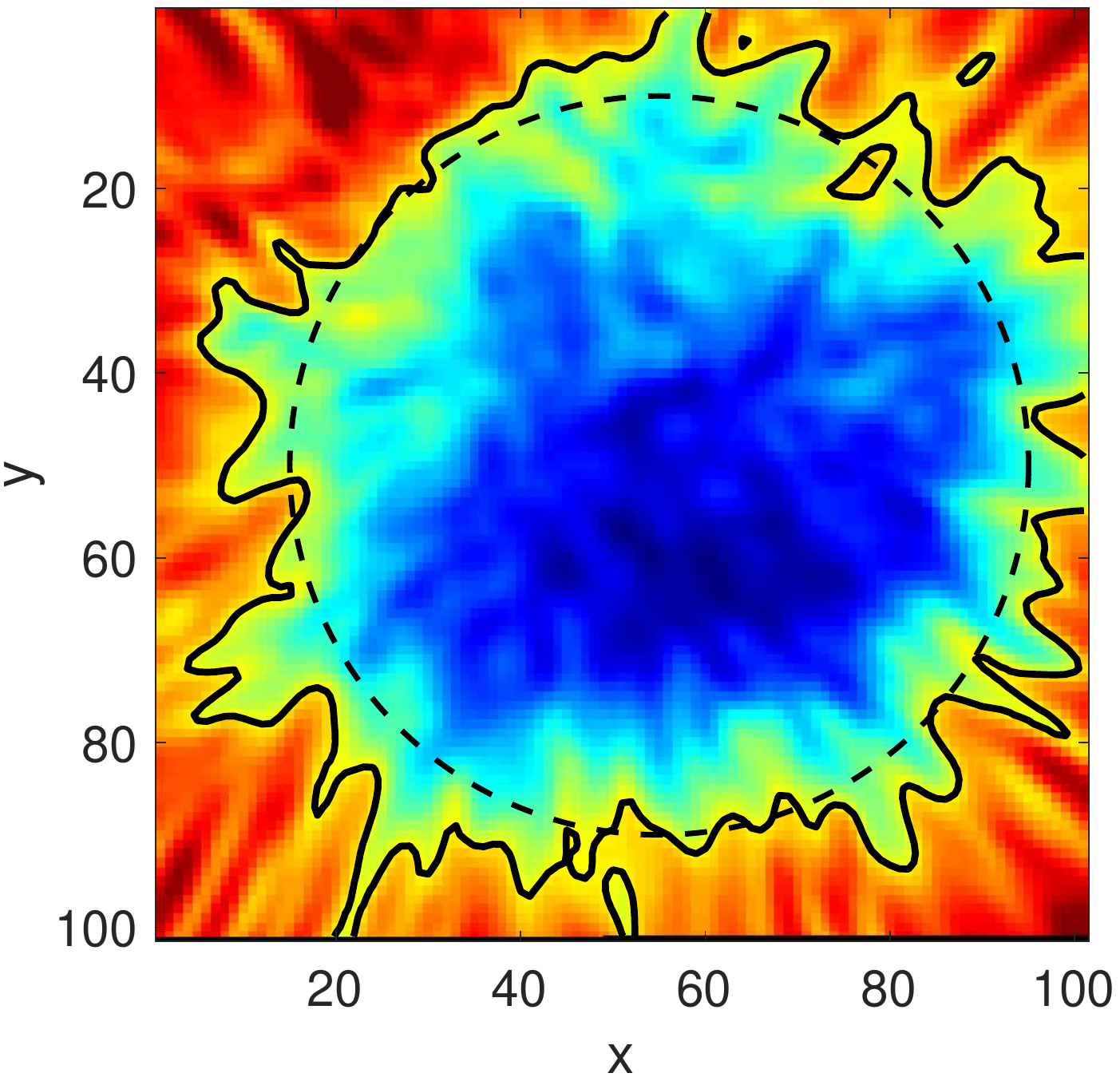}&
    
    \includegraphics[width=40mm]{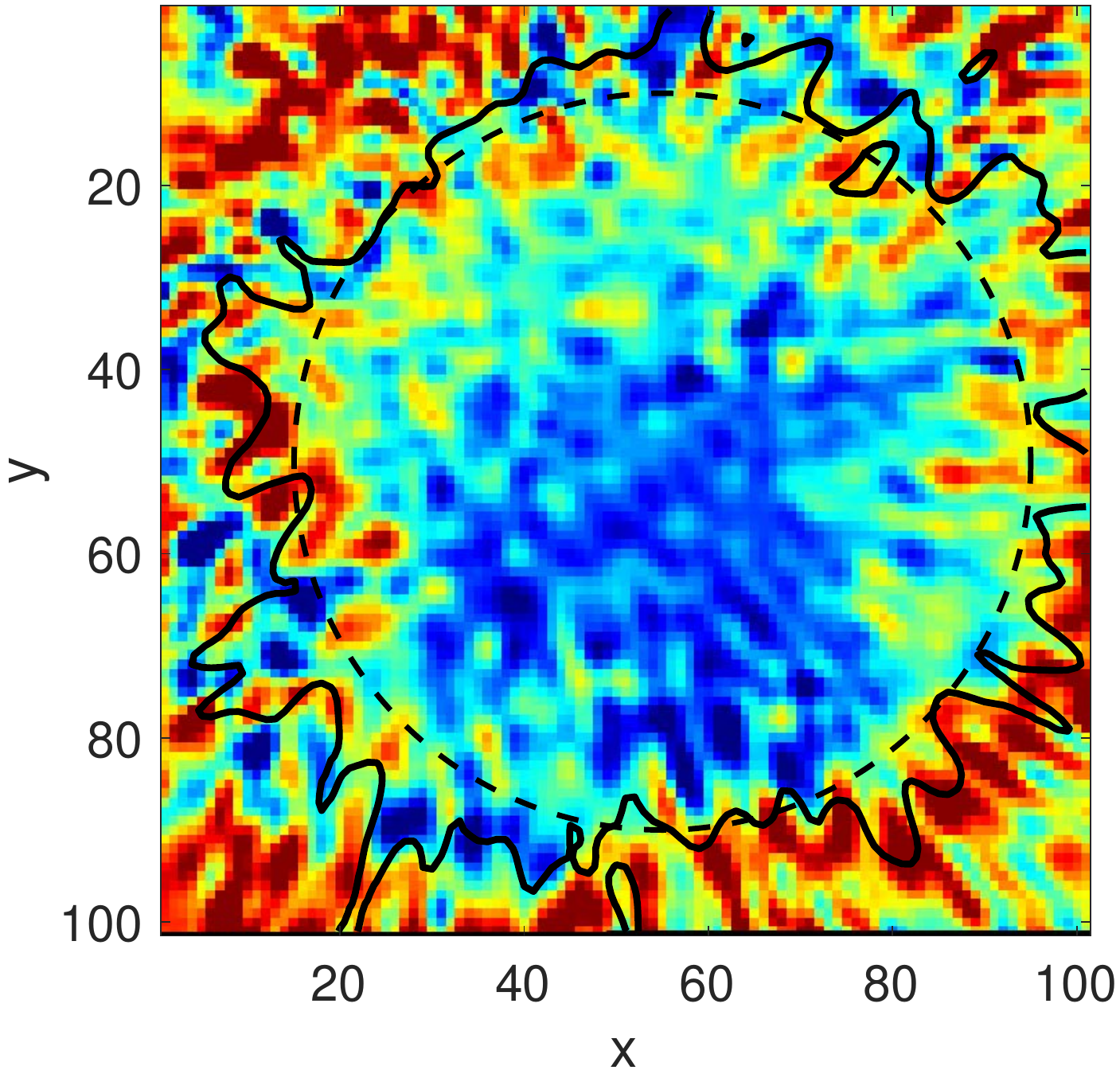}&

     \includegraphics[width=45mm]{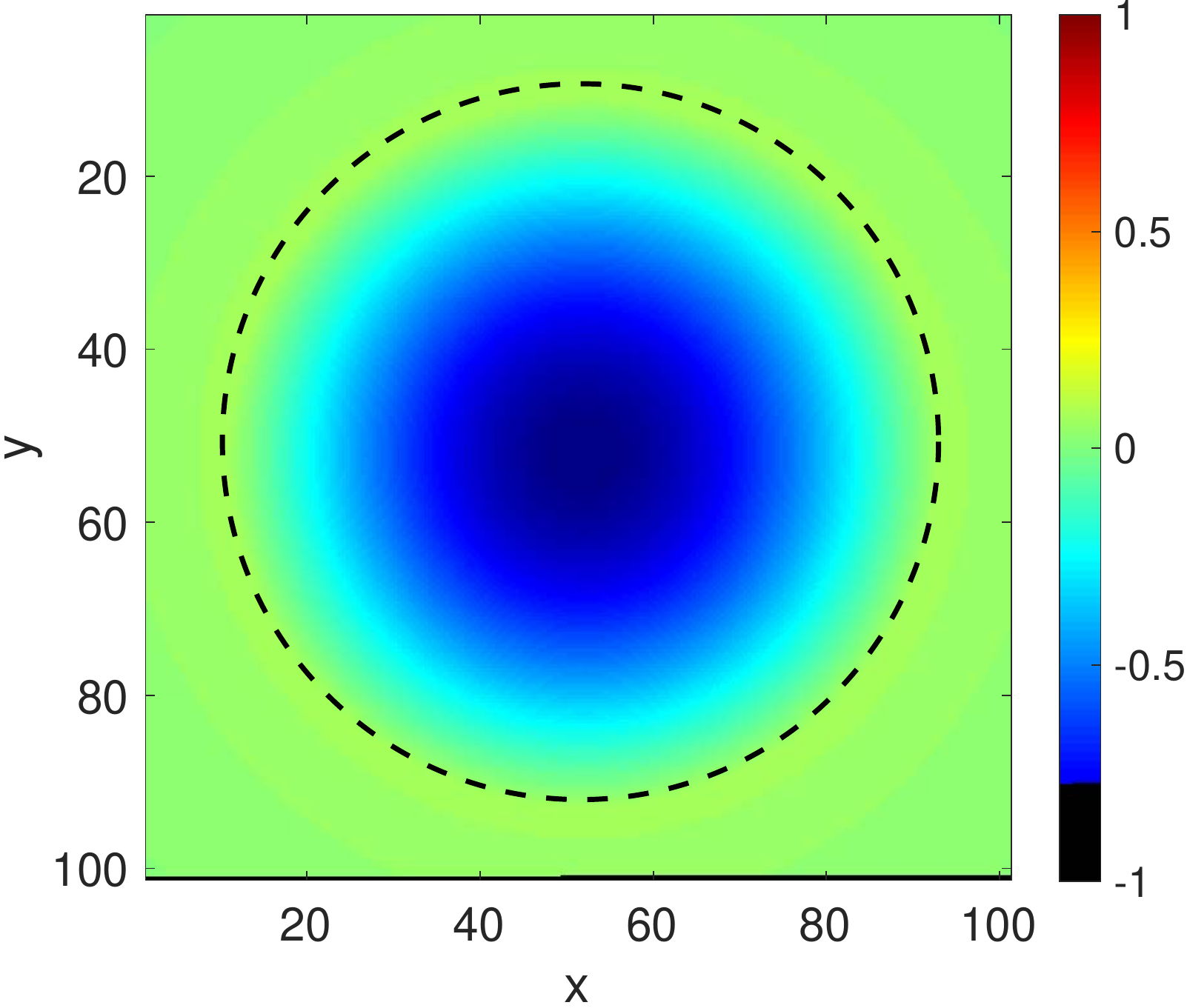}\\

  \end{tabular}
\caption{The first column displays the spatial structure of the first POD mode with peak power at $f=4.9$ mHz. The second column displays the spatial structure of the DMD mode that corresponds to the same frequency of $f=4.8$ mHz. The third column shows the density perturbation of a slow sausage body mode in a cylindrical magnetic flux tube and the dashed circle shows the boundary. In the first and the second columns the solid black line shows the umbra/penumbra boundary as shown in the middle panel of Figure \ref{fig:POD} and the dashed circle is used to compare the observations with the flux tube in the third column. The images shown in the two rows are chosen to be in anti-phase, hence, they represent different time snapshots.
\label{fig:Susage}}. 
\end{figure}

\begin{figure}[htp]
  \centering
  \begin{tabular}{ccc}
    
     \includegraphics[width=40mm]{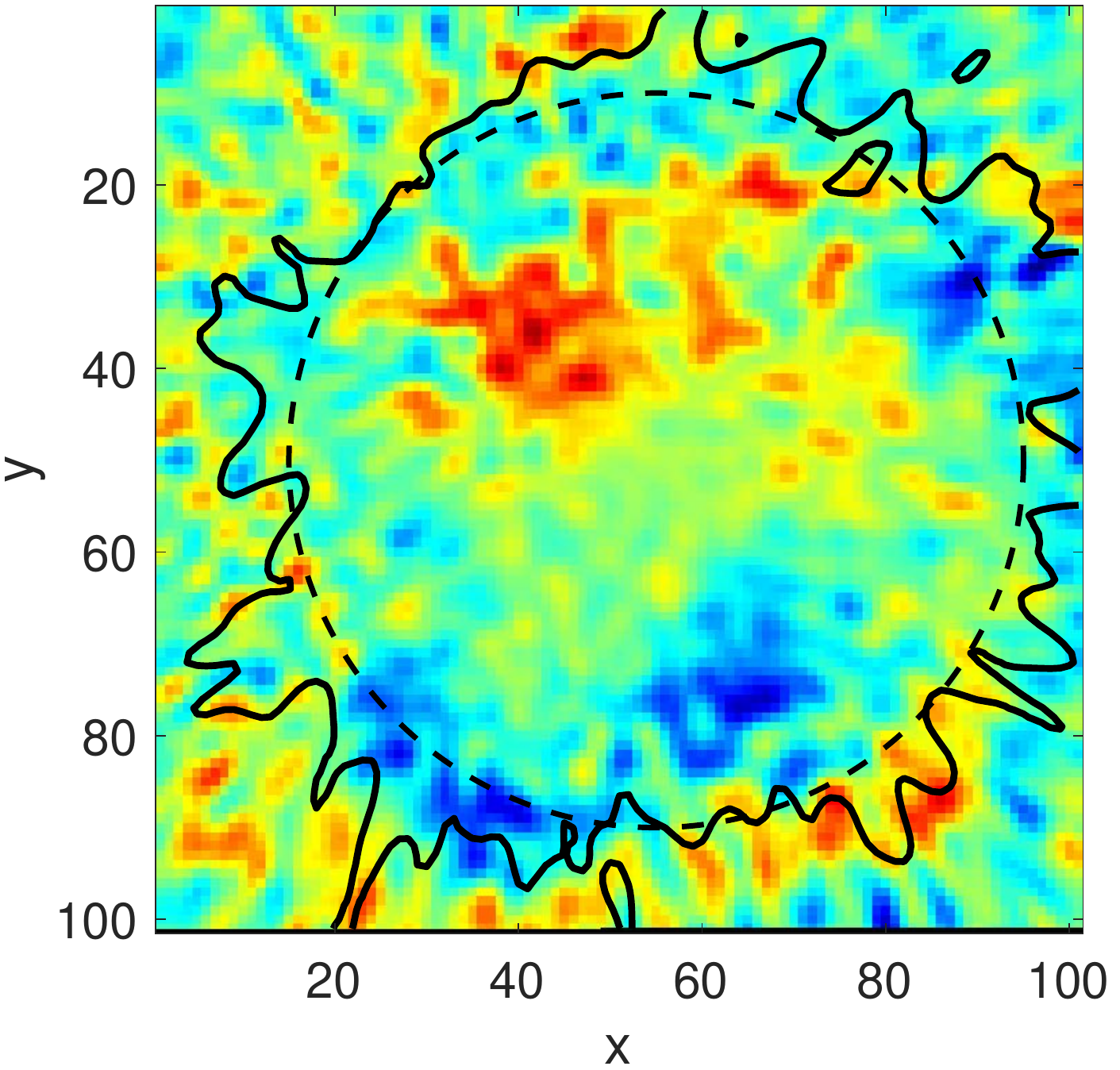}&
     
     \includegraphics[width=40mm]{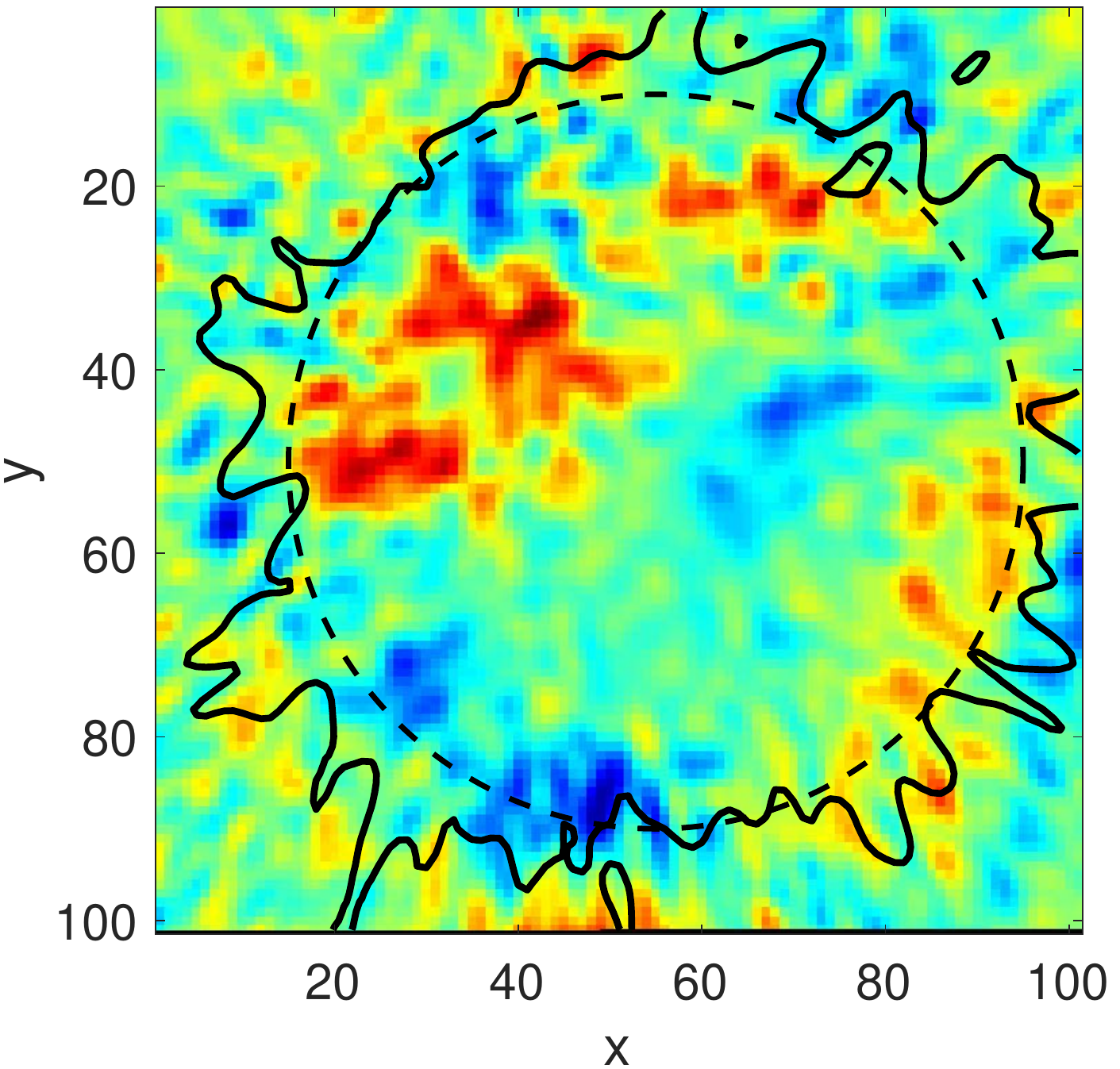}&

     \includegraphics[width=45mm]{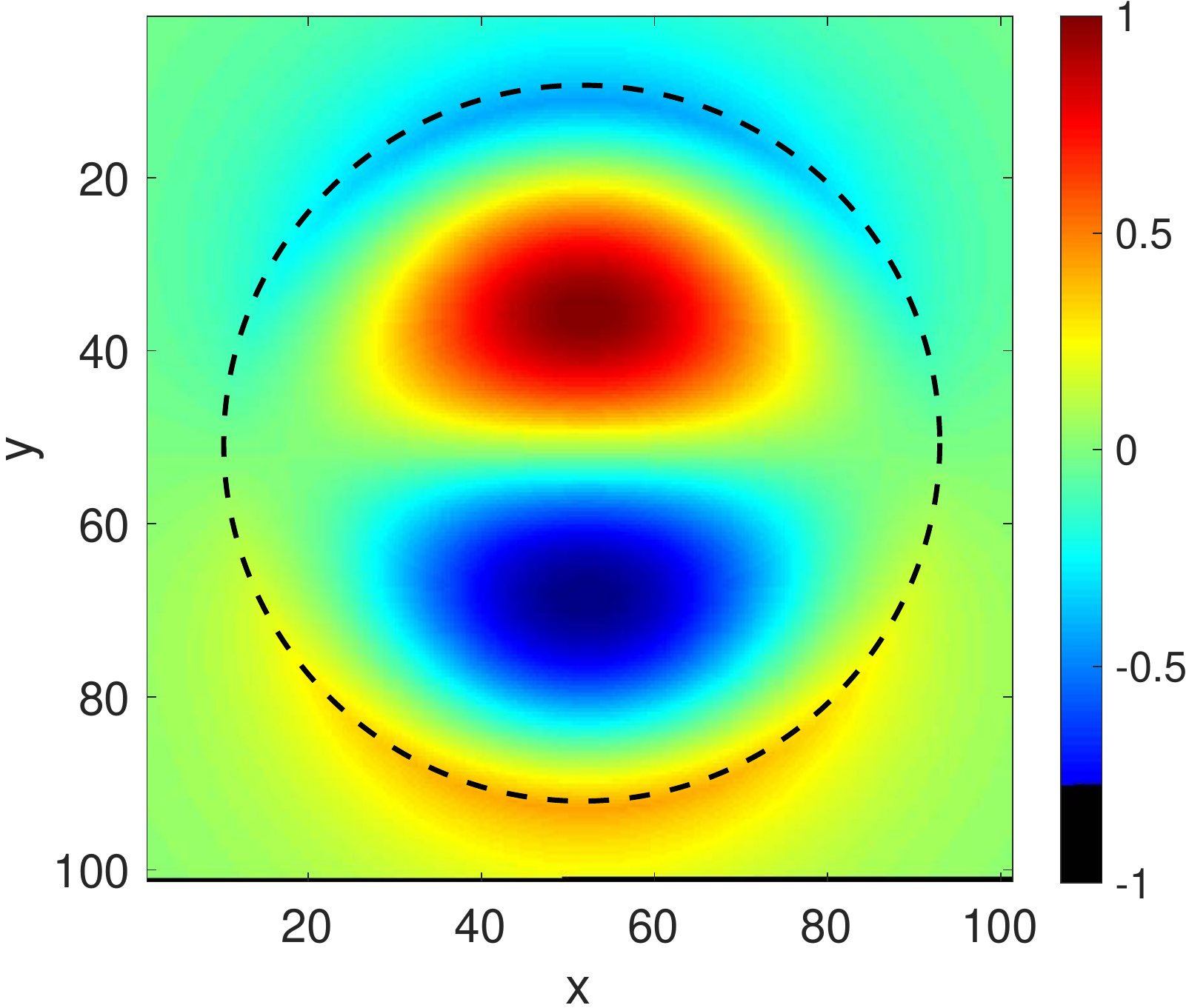}\\
     
    \includegraphics[width=40mm]{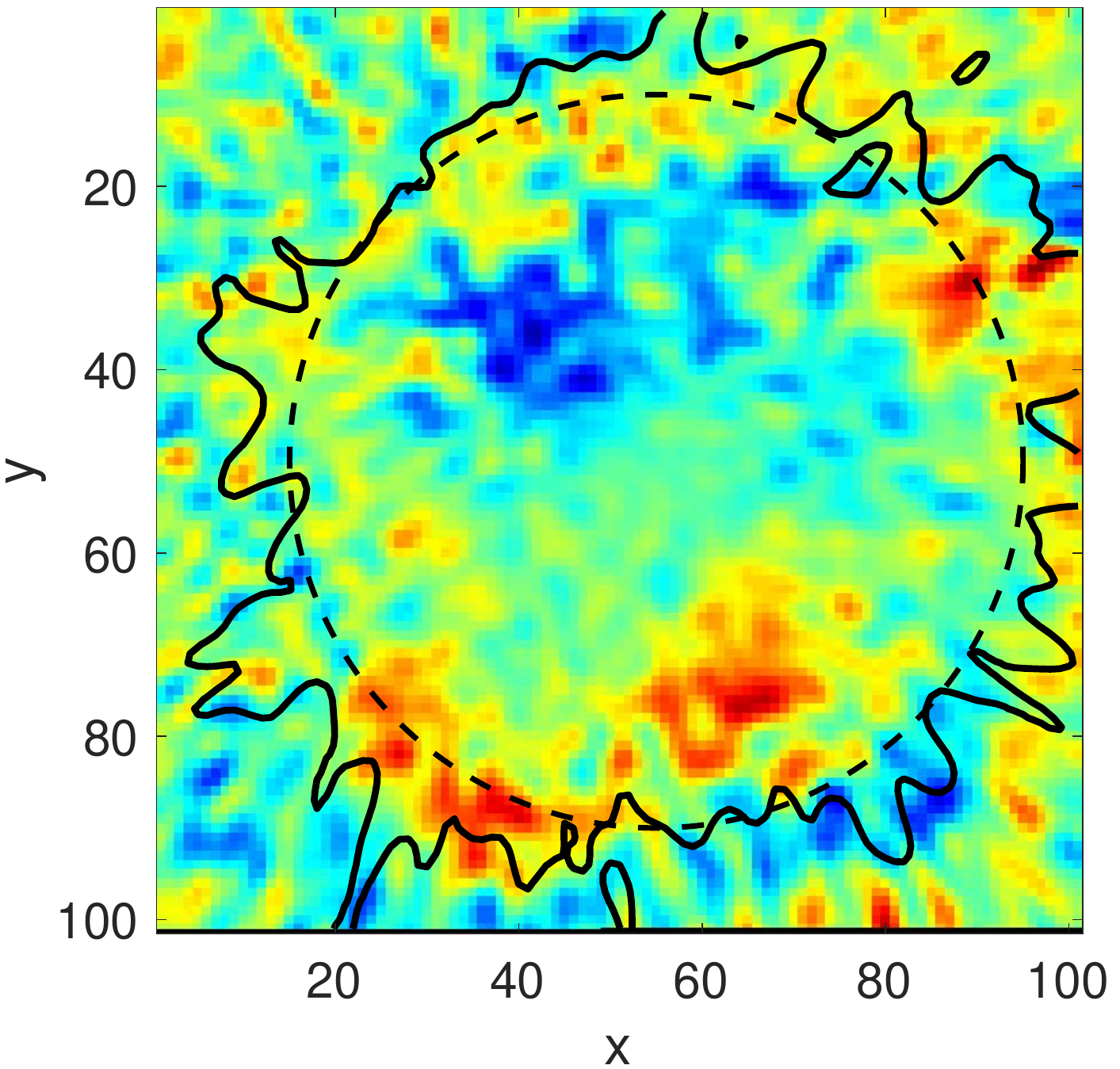}&
    
     \includegraphics[width=40mm]{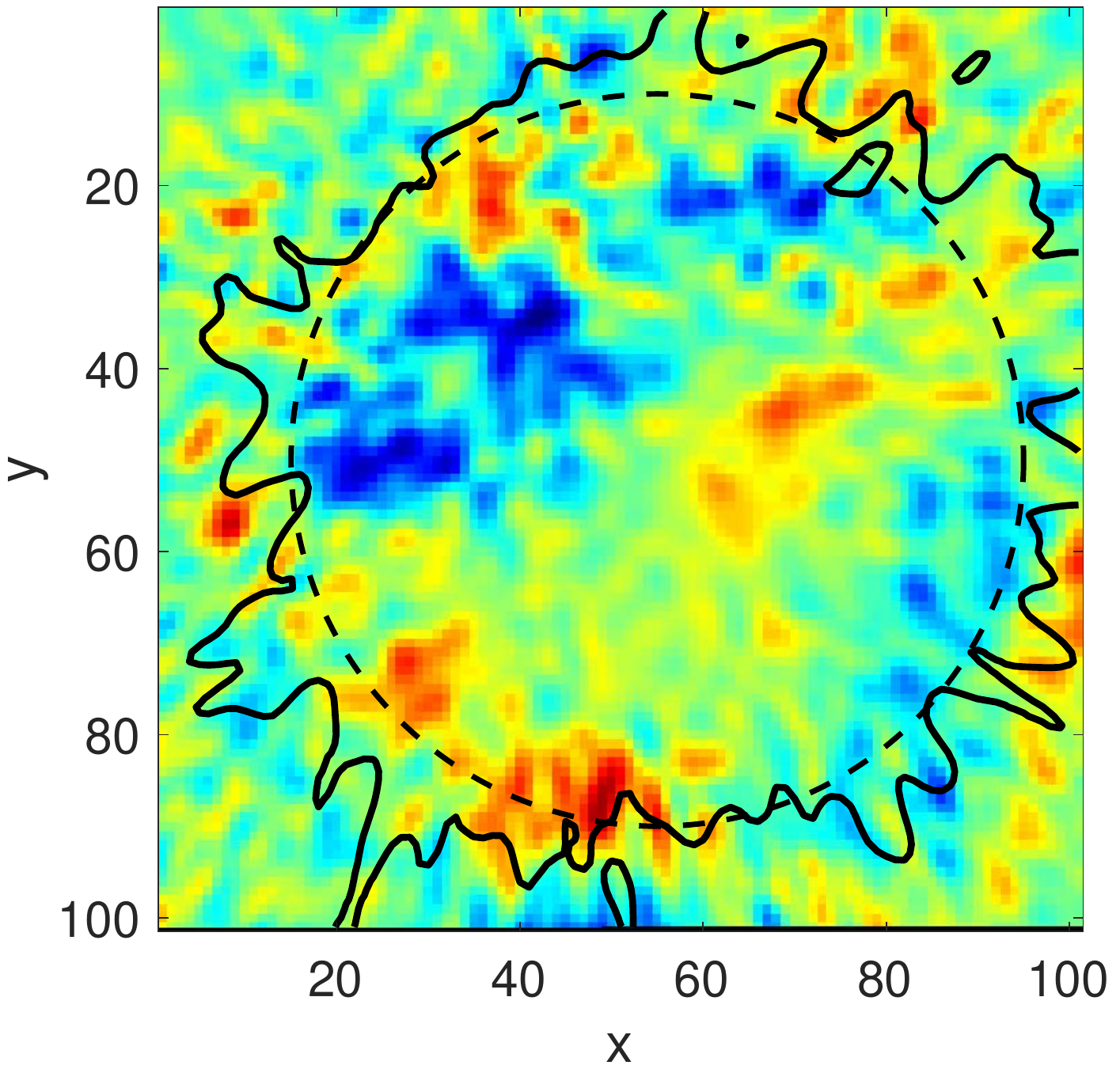}&

     \includegraphics[width=45mm]{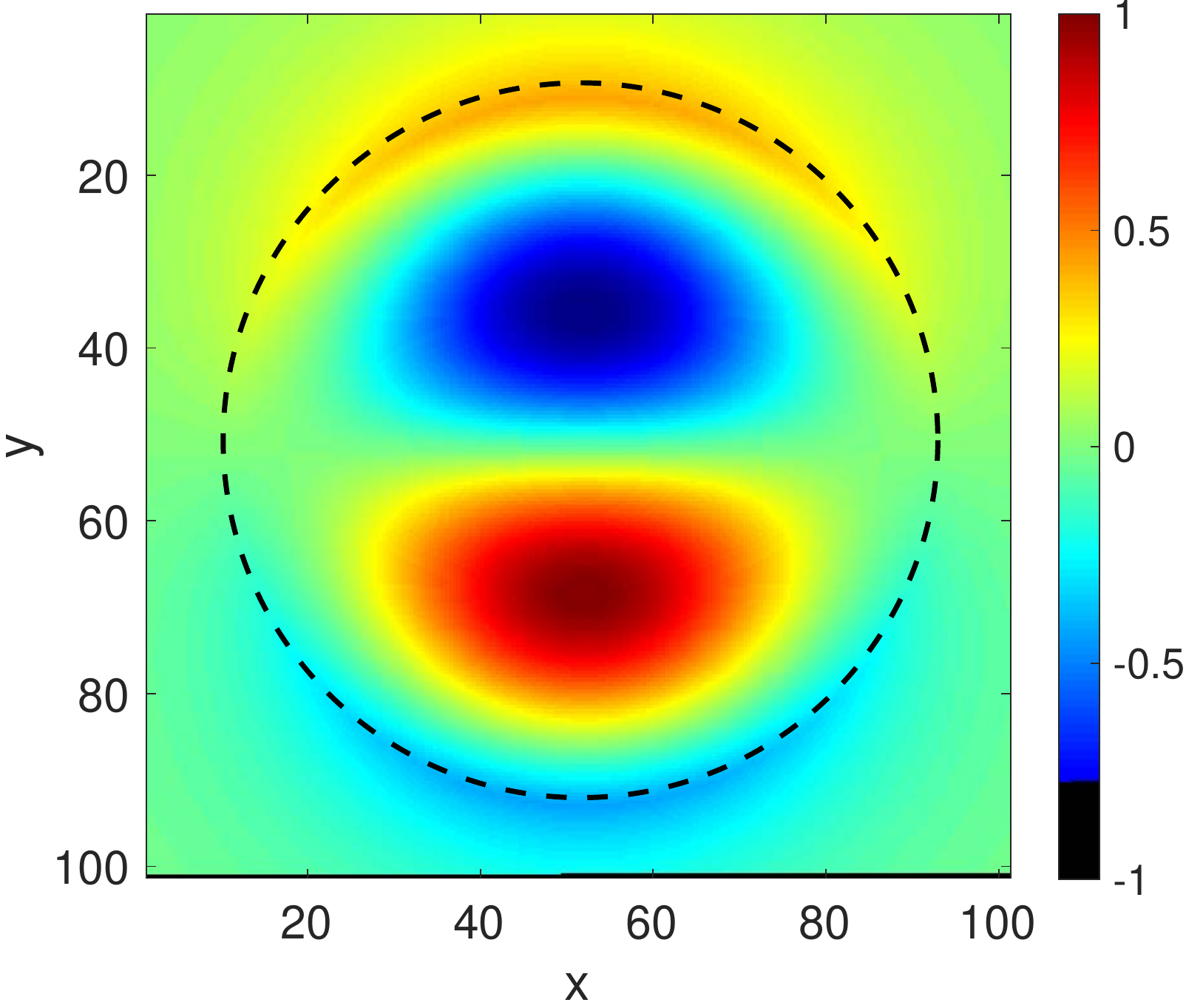}\\
     
  \end{tabular}

\caption{The first column displays the spatial structure of the $13^{th}$ POD mode with peak power at $f=6$ mHz. The second column displays the spatial structure of the DMD mode that corresponds to the same frequency of $f=6$ mHz. The third column shows the density perturbation of a slow kink body mode in a cylindrical magnetic flux tube and the dashed circle shows the boundary of the tube. In the first and the second columns the solid black line shows the umbra/penumbra boundary as shown in the middle panel of Figure \ref{fig:POD} and the dashed circle is to compare the observations with the flux tube in the third column. The images shown in the two rows are chosen to be in anti-phase, hence, they represent different time snapshots.
\label{fig:kink}}.  
\end{figure}

\begin{figure}[htp]
  \centering
  \begin{tabular}{cc}
    
     \includegraphics[width=60mm]{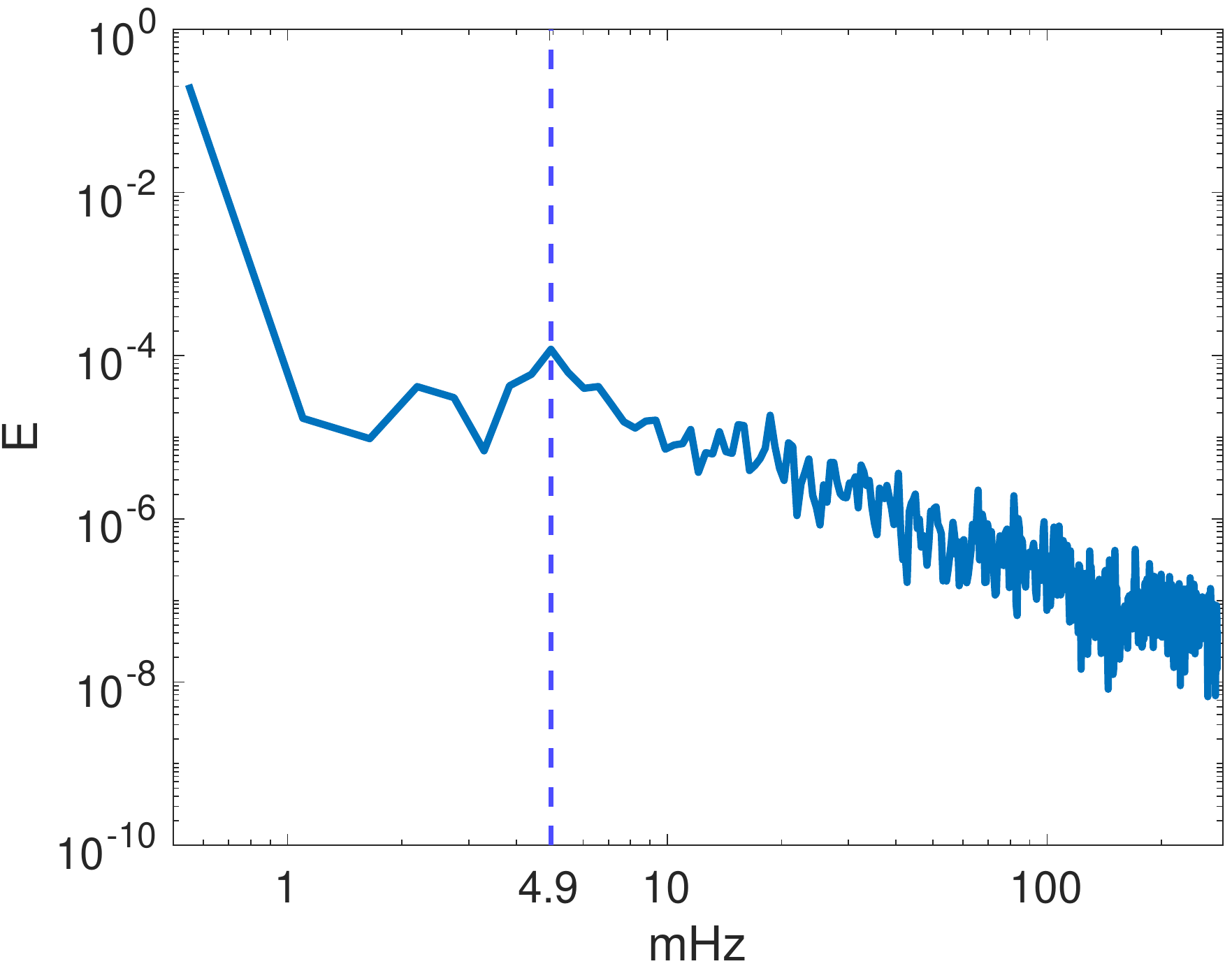}&

     \includegraphics[width=60mm]{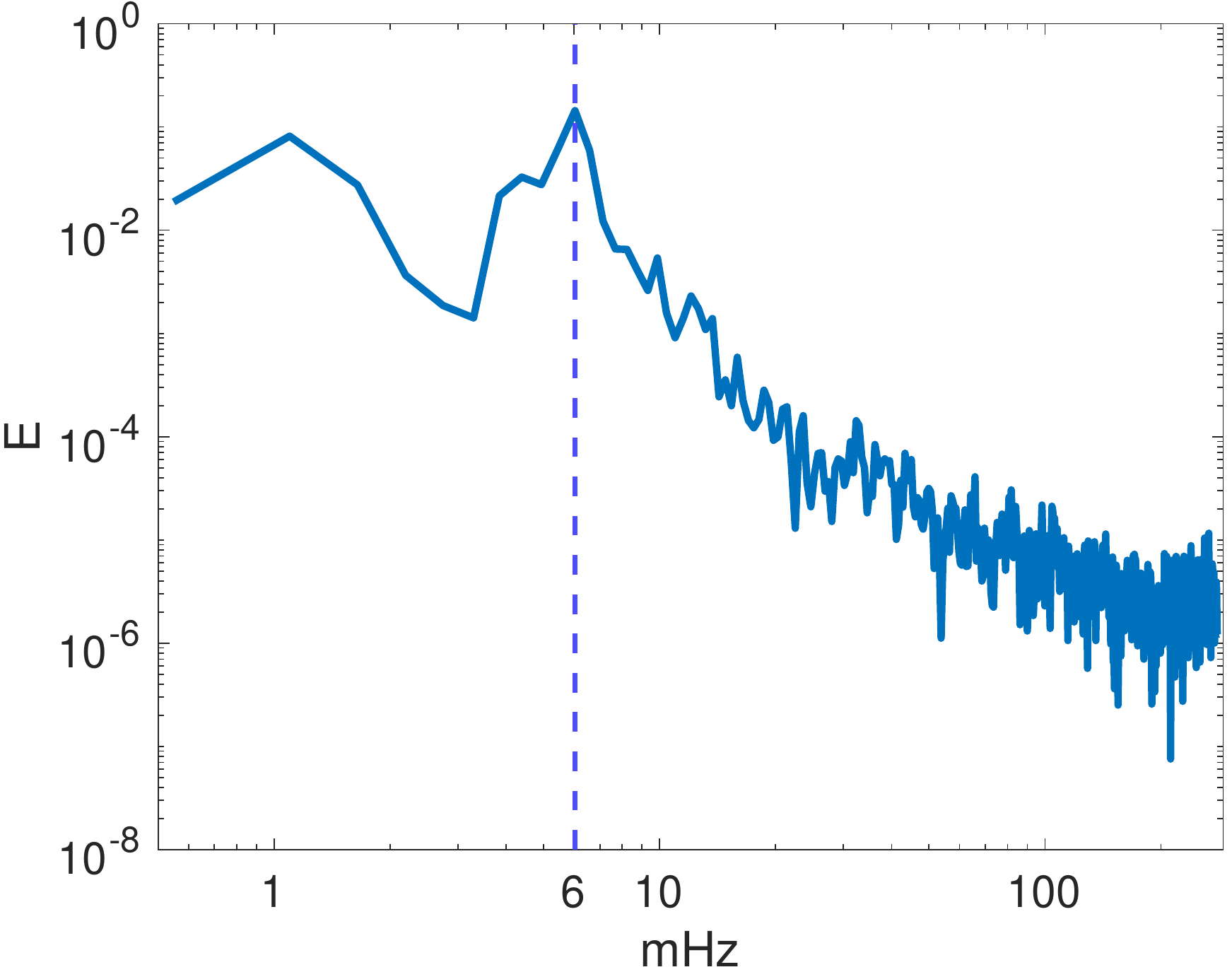}\\

  \end{tabular}

\caption{The left panel displays the PSD of POD 1 mode and it has a peak at $4.9$ mHz, while the right panel displays the PSD of POD 13 mode showing its a peak at $6$ mHz.
\label{fig:PSD1&PSD13}}
 \end{figure}

\subsection{Dynamic Mode Decomposition (DMD)}

The DMD technique, first introduced by Schmid \cite{schmid2010dynamic}, is a data-driven algorithm that can extract the dynamic information of the flow generated by numerical simulations or in a measured physical experiment \cite{hemati2014dynamic}. DMD modes represent the spatial structure of the mode where the associated eigenvalues give information about the oscillation frequencies of the modes. DMD is a widely used technique in the field of fluid mechanics, e.g., jet flows \cite{rowley2009spectral, jovanovic2014sparsity}, bluff body flows \cite{bagheri2013koopman} and visco-elastic fluid flows \cite{grilli2013transition}). It can therefore also extract information about the coherent spatial structure of observed MHD wave modes if the modes have distinct frequencies as we show in this study.

To apply the DMD technique the time snapshots have to be organized in columns analogously to POD, but in two matrices defined as
\begin{equation}
    W^{A}=\{w_{1},w_{2},...w_{\tau} \} \ \ \  and \ \ \ W^{B}=\{w_{2},w_{3},...w_{T} \},
\end{equation}
where $\boldmath{W^{B}}$ is shifted by a snapshot of $\boldmath{W^{A}}$ such that $\tau=(T-1)$. The  matrices $W^{A}$ and $ W^{B}$ are related by a linear operator $A \in \mathbb{C}^{N\times N}$ as 
\begin{equation}
    W^{B}=A W^{A}. \label{eq4}
\end{equation}
DMD is based on approximating the eigenvalues and eigenvectors of the linear operator $A$ without actually computing them exactly since for most practical applications the size of $A$ is too large. Using SVD the matrix $\boldmath{W^{A}}$ is decomposed as
\begin{equation}{}
     W^{A}=\tilde{\Phi}\tilde{S}\tilde{C^{*}}
\end{equation}{}
 and substituted in Eq. (\ref{eq4}) to give
 \begin{equation}
     \tilde{\Phi}^{*}W^{B}\tilde{C}\tilde{S}^{-1}=\tilde{\Phi}^{*} A  \tilde{\Phi}.
\end{equation}
From this we define
\begin{equation}
    \tilde{A}=\tilde{\Phi}^{*} A \tilde{\Phi},
\end{equation}
where $\tilde{\boldmath{A}} \in \mathbb{C}^{\tau \times \tau}$ is the optimal low-dimensional representation of $A$, (note that $\tau\ll N$) so that we can calculate the complex eigenvalues, $\mu_{i}$, and associated eigenvectors, $z_{i}$, of $\tilde{\boldmath{A}}$, where $i=1...\tau$. 

To obtain the spatial structure of the DMD modes, we follow the method developed by Jovanovic et al. \cite{jovanovic2014sparsity} by calculating a Vandermonde matrix,
\begin{equation}
   Q_{i,j}=\mu^{j-1}_i, 
\end{equation}
where $i=1...\tau$ and $j=1...\tau$. After this operation is completed, the spatial structure of the DMD modes are obtained from \begin{equation}
    \Psi=W^{A}Q^{*},
\end{equation}{}
and the distinct frequencies associated with each these modes are
\begin{equation}
   f_i=f_s \mathrm{arg}(z_i)/2\pi, 
\end{equation}
where $f_s$ is the sampling frequency.

\section{Method, results and MHD wave mode identification} \label{sec:resl}

Our goal is to use POD and DMD in combination to identify coherent oscillations across the sunspot's umbra and compare these modes with the MHD wave modes of a cylindrical magnetic flux tube predicted from theory.

The particular ROI of the sunspot umbra to be studied is shown by the green box in Figure \ref{fig:POD}. Firstly, this ROI is analysed using the POD technique, which ranks modes based on their contribution to the overall variance. This step is followed by the calculation of the power spectral density (PSD) of the POD time coefficient associated with each of these modes. The PSD of the first 20 modes, which contains the majority of the energy (96.86 \%), show frequency peaks between $4.3$ mHz and $6.5$ mHz as shown in the right panel of Figure \ref{fig:POD}. The PSD of the individual POD modes are then used to determine the dominant frequency, or frequencies if there are a mix of frequencies, associated with a particular POD mode, so that this information could be applied to determine the coherent spatial structure of modes with distinct frequencies using DMD. If there is no exact match between frequencies, the DMD mode closest to the target frequency is selected. 

For illustrative purposes we will concentrate on the first branches of the sausage and kink slow body modes, i.e. modes with only one radial node occurring at the umbra/penumbra boundary. The first POD mode shows the clear azimuthal symmetry of a sausage mode as shown in the first column in Figure \ref{fig:Susage}, with a PSD peak at $4.9$ mHz as shown in the left panel of Figure \ref{fig:PSD1&PSD13}. The DMD mode that corresponds to the same frequency of $4.8$ mHz is shown in the second column in Figure \ref{fig:Susage}. The third column shows the density perturbation of the slow body sausage mode from the cylindrical magnetic flux tube model. This is important for comparison since the MHD wave modes in a cylindrical flux tube are, by definition, spatially orthogonal. Since the observed umbra is approximately circular, POD, which defines modes by spatial orthogonality, should perform well in this particular case study. What is more remarkable is that the DMD technique, which does not have any such criteria, still manages to identify the sausage mode. From both the POD and DMD analysis there is strong oscillatory power in the penumbra at $4.8$ mHz and the penumbral filaments can cleary be identified. Obviously, the idealised cylindrical magnetic flux tube model cannot recreate this oscillatory behaviour since it assumes a simple quiescent environment without complex fibril structuring. In addition, it is important to state that even within the umbra, disagreement between observations and the eigenmodes of a magnetic cylinder could simply be due to the fact that the observed oscillations are being continually forced and are not free. 

The next POD mode that can be interpreted as a physical MHD wave mode is the $13^{th}$ mode which has the azimuthal asymmetry of a kink mode as shown in the first column of Figure \ref{fig:kink}, with a peak at $6$ mHz as shown in the right panel of Figure \ref{fig:PSD1&PSD13}. The DMD mode with frequency of $6$ mHz is shown in the second column in Figure \ref{fig:kink}. Again, for comparison the slow kink body mode from cylindrical flux tube theory is shown in the third column.  Here we can compare these results with the previous work of J17. These authors identified a kink mode rotating in the azimuthal direction by implementing a $k-\omega$ Fourier filter ($0.45 - 0.90$ arcsec$ ^{-1}$ and $5 - 6.3$ mHz). Hence, the kink mode frequency from POD and DMD is certainly in the same frequency range as the filter applied by J17. Our analysis reveals that the time coefficients of the POD modes are almost sinusoidal. This is remarkable since POD puts no such condition on these coefficients. Hence, Fourier analysis, which has a sinusoidal basis in the temporal domain, in retrospect was a valid approach. The problem with Fourier analysis is the assumption of a sinusoidal basis in the  spatial domain, since in the cylinder model, the basis functions in the radial direction are Bessel functions. The strength of POD is that it calculates a spatially orthogonal basis, regardless of the geometry of the observed waveguide. Also, the further advantage of both POD and DMD over Fourier analysis is that these methods cross-correlate individual pixels in the ROI, in the spatial and temporal domain, respectively. This is a distinct advantage in identifying a coherent oscillations across the whole umbra. In agreement with the sausage mode identification in Figure \ref{fig:Susage}, the spatial structure of the POD and DMD modes in the first and second columns of Figure \ref{fig:kink} is very similar even though the DMD places no restriction on the mode being orthogonal. This further strengthens the argument that the kink mode interpretation is indeed physical.

\begin{figure}[htp]
  \centering
  \begin{tabular}{cc}
    
     \includegraphics[width=50mm]{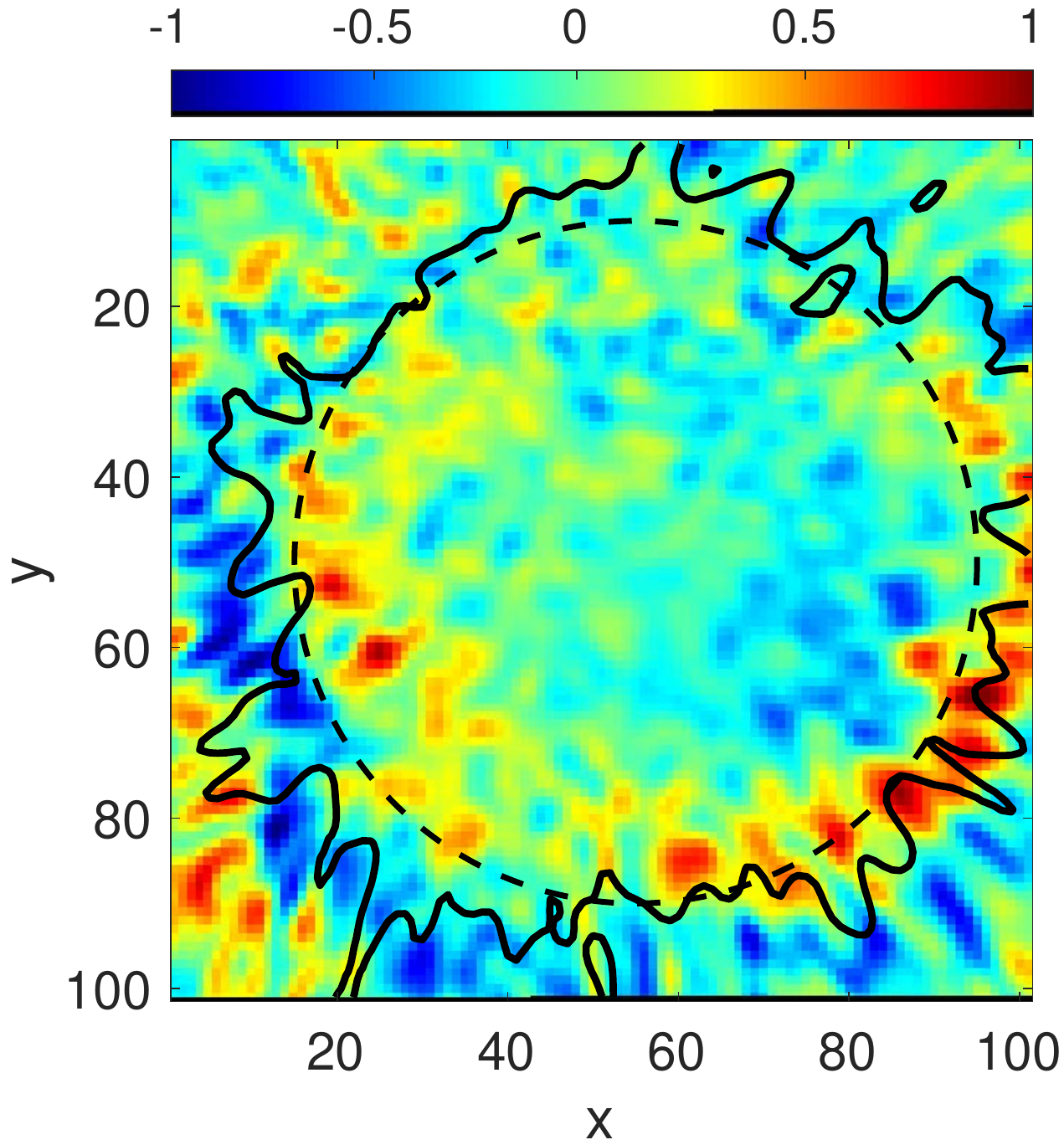}&

    \includegraphics[width=50mm]{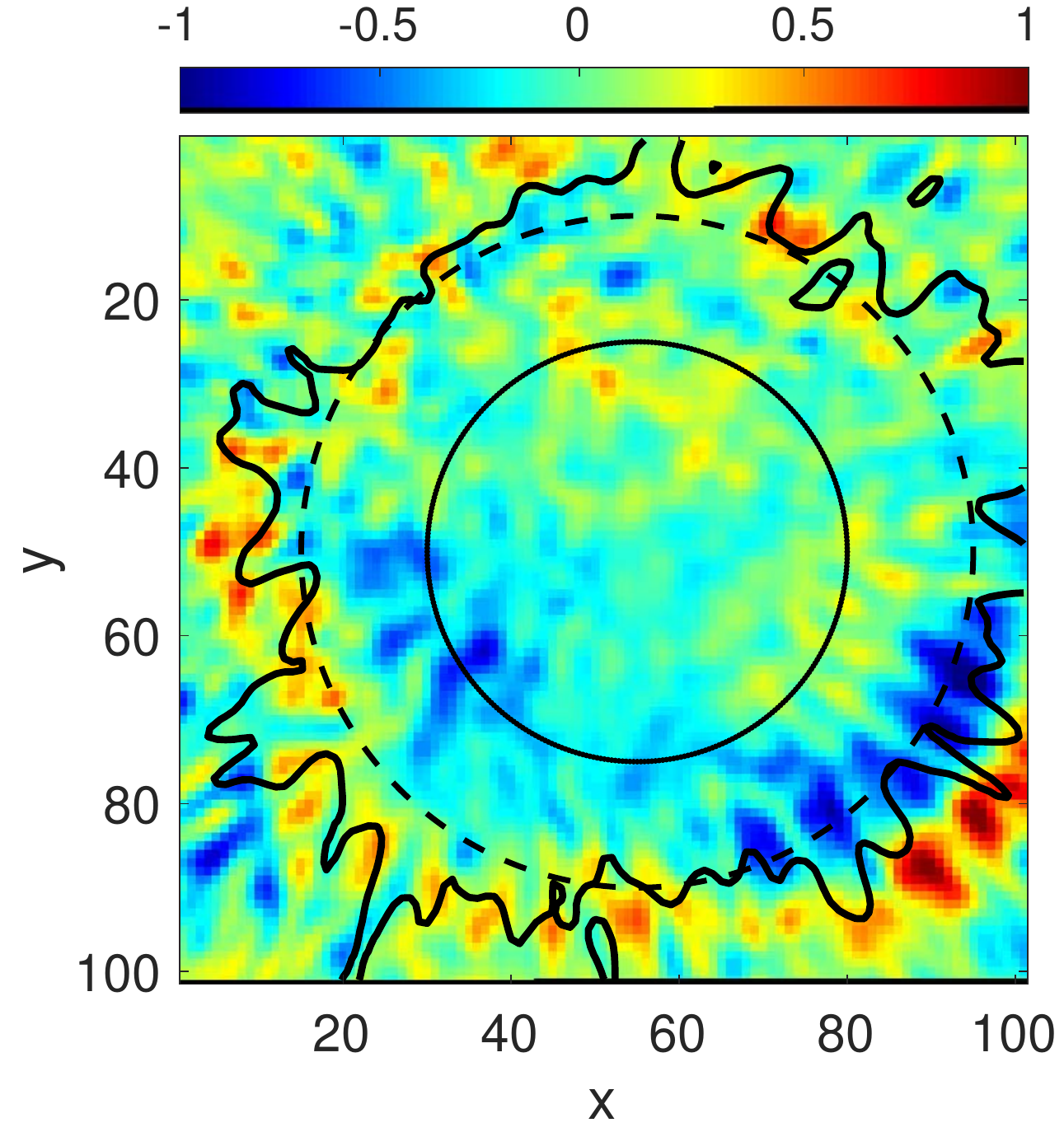}\\
 \end{tabular}
  
\caption{Left panel displays POD 10, which is orthogonal in space to POD 13 which is shown on the first column of Figure \ref{fig:kink}. The right panel shows the DMD mode with a frequency of $5.4$ mHz and it is approximately orthogonal in space to the DMD mode with a frequency of $6$ mHz displayed in the second column of Figure \ref{fig:kink}. The solid black circle shows the path of the time-azimuth diagram in Figure \ref{fig:T_D}. \label{fig:POD10}}
 \end{figure}

\begin{figure}[htp]
  \centering
  \begin{tabular}{cc}
    
     \includegraphics[width=60mm]{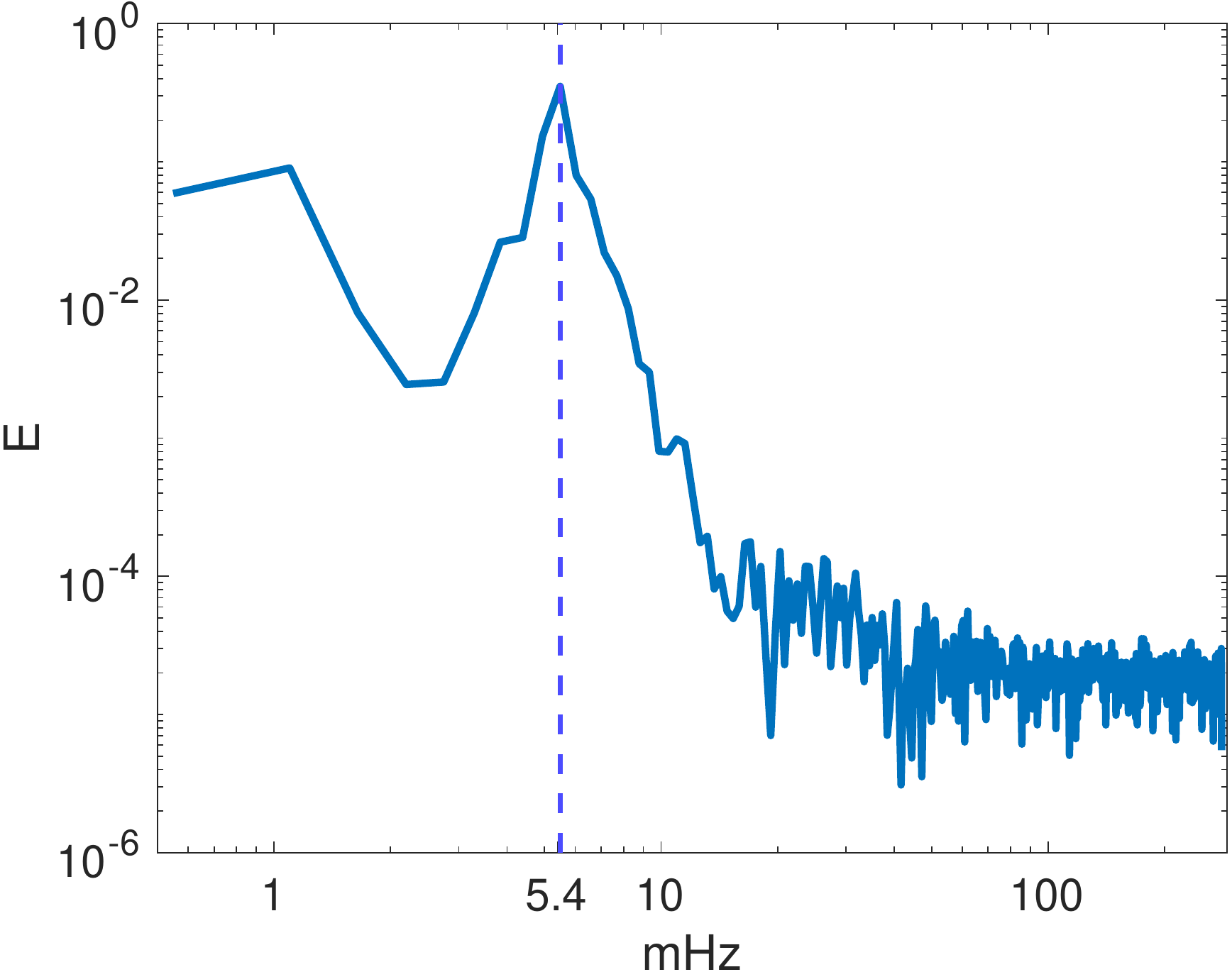}&

     \includegraphics[width=60mm]{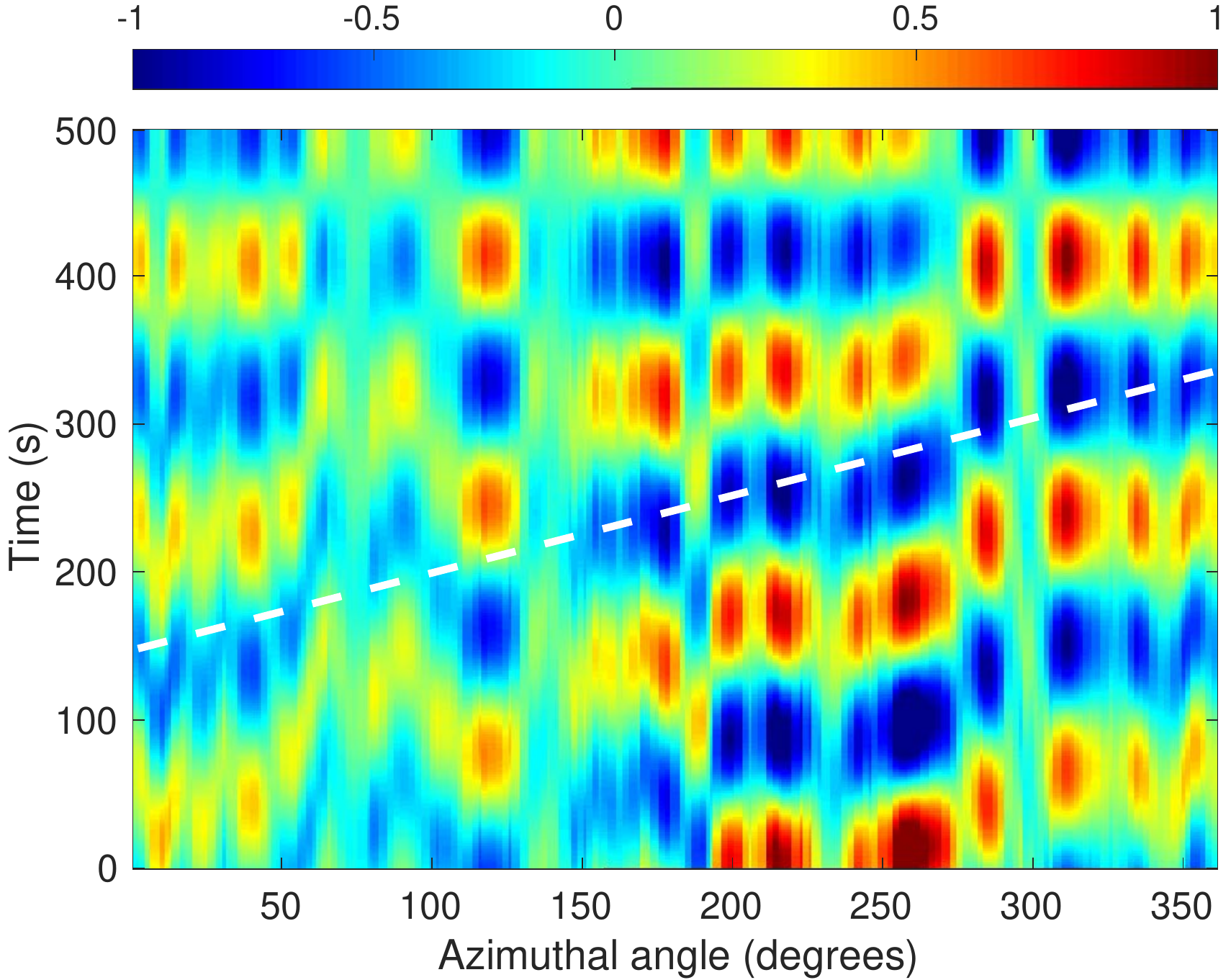}\\

  \end{tabular}

\caption{The left panel displays the PSD of POD 10 and it has a peak at $5.4$ mHz. The right panel shows the time-azimuth diagram after the superposition of two approximately spatially perpendicular kink modes identified with DMD. The white dashed line on the right panel shows gives an apparent angular velocity  of about $2\deg$ s$^{-1}$ consistent with the result of J17 \label{fig:T_D}}
 \end{figure}

Here we would like to investigate the apparent rotational motion of the kink mode detected by J17 who constructed a time-azimuth diagram around the circumference of the umbra and estimated an angular velocity of approximately $2 \deg$ s$^{-1}$ and a periodicity of about 170 s. Physically, the rotational motion could be explained by having either (i) a kink mode that is standing in the radial direction but propagating in the azimuthal direction or (ii) it could be the superpostion of two approximately perpendicular kink modes (both standing in the radial and azimuthal directions). Before attempting to recover this rotational motion with the POD and DMD techniques it should be emphasised that the filtering process performed by J17 crudely oversimplified the complexity of the swirling "washing machine" motion in the original signal. In particular, the 40 s wide temporal filter could contain at least least 7 DMD modes. Spatially, the filter effectively divided the umbra into quadrants. To recreate the apparent rotational motion (or approximate circular polarisation) with POD we need to superimpose at least two spatially perpendicular kink modes with similar, but not necessarily identical, periods. From our analysis this requires the superposition of POD 10 (shown on the left panel of Figure \ref{fig:POD10} and POD 13 shown on the first column on Figure \ref{fig:kink}). The PSD of POD 10 has a peak at $5.4$ mHz as shown on the left panel of Figure \ref{fig:T_D}, while PSD of POD 13 has a peak at $6$ mHz as shown on the right panel of Figure \ref{fig:PSD1&PSD13}. Both these frequencies lie within the temporal filter chosen by J17. We can also recreate this rotational motion by superimposing at least two DMD modes. Although DMD modes are not defined to be orthogonal in space, we still find two examples of kink modes with DMD that are approximately perpendicular to each other and are also in the same frequency range of J17. These modes correspond to a frequency of $5.4$ mHz (see the right panel of Figure \ref{fig:POD10}) and $6$ mHz shown on the second column of Figure \ref{fig:kink}. A similar time-azimuth analysis to J17 was performed on the superposition of these two DMD modes along the solid black circle shown on the right panel of Figure \ref{fig:POD10} where the signal was strongest. This resulted in an angular velocity of about $2 \deg$ s$^{-1}$ and periodicity of approximately 170 s (see the right panel of Figure \ref{fig:T_D}), consistent with the result of J17.

\begin{figure}[htp]
  \centering
  \begin{tabular}{cc}
    
     \includegraphics[width=50mm]{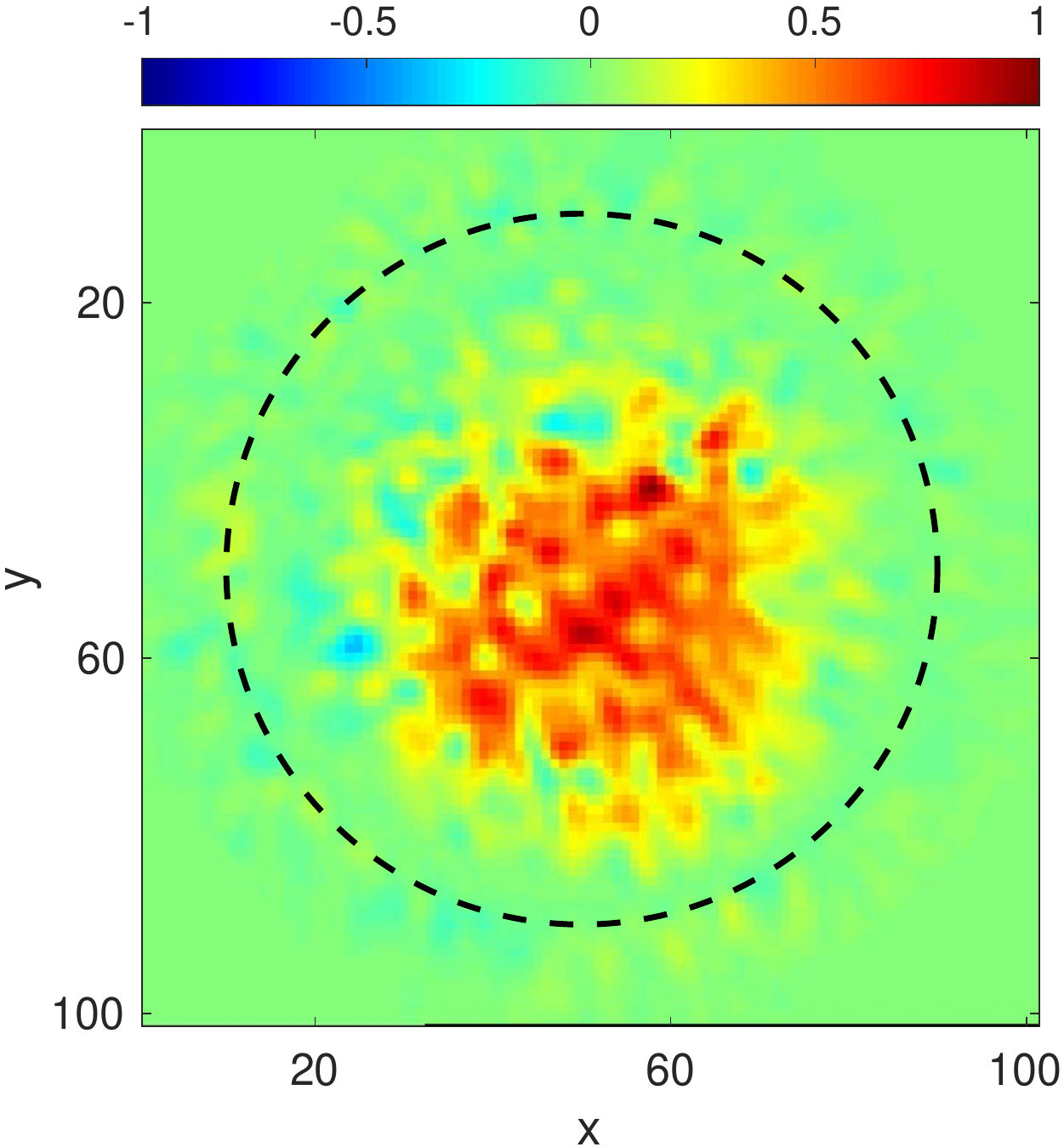}&

     \includegraphics[width=50mm]{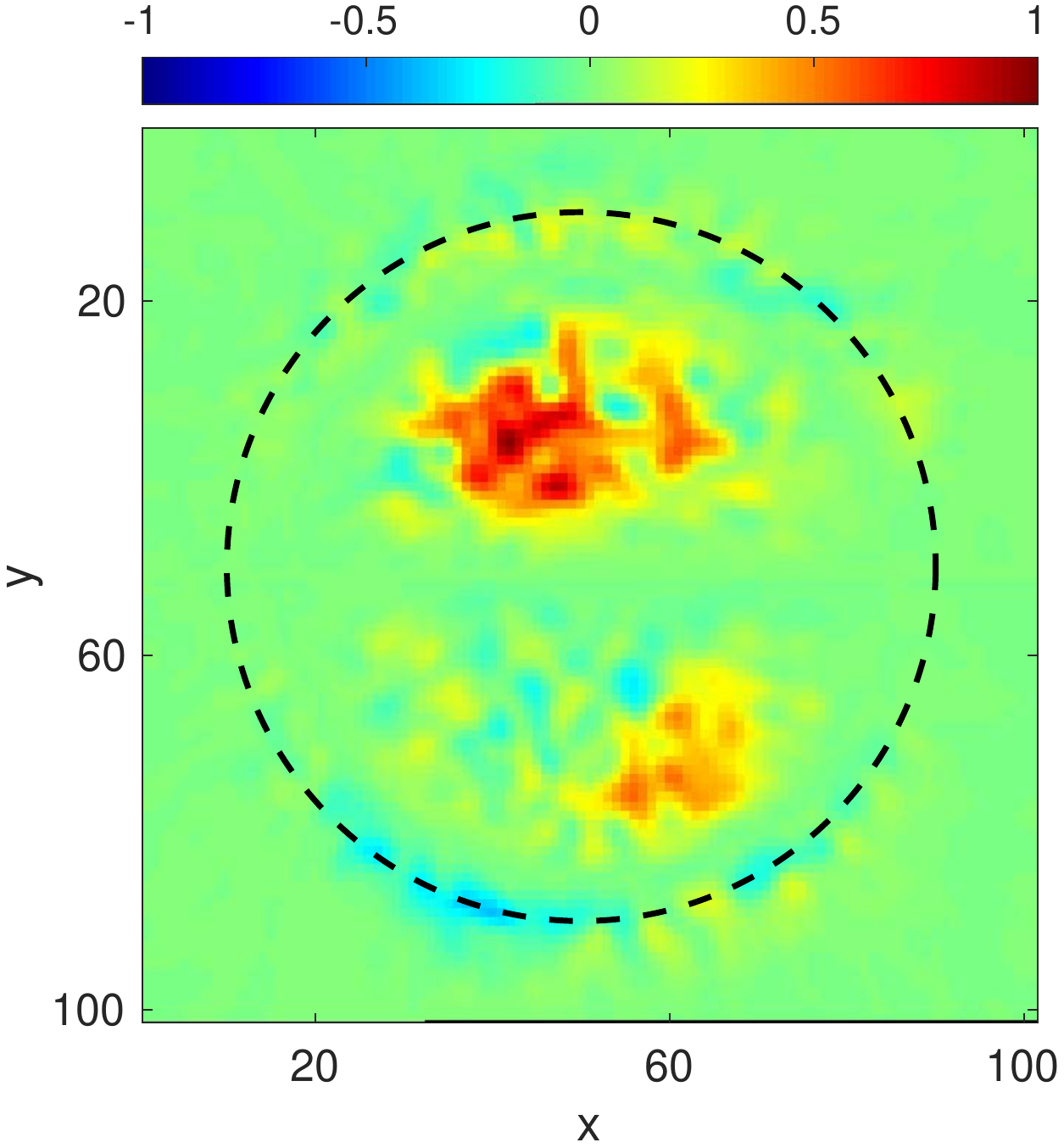}\\

  \end{tabular}

\caption{The left panel displays the Pearson correlation between theoretically constructed and observationally detected sausage mode shown in Figure \ref{fig:Susage} and the kink mode shown in Figure \ref{fig:kink}. The positive/negative numbers in the colourbar denote correlation/anti-correlation.  \label{corr}}
 \end{figure}

To compare the cylinder model MHD modes with the POD modes from the observational data, we also performed a Pearson correlation analysis, calculated on a pixel-by-pixel basis for the sausage (see Figure \ref{fig:Susage}) and kink (see Figure \ref{fig:kink}) modes using as shown on Figure \ref{corr}. The result of the correlation is a number between 1 and -1, where 1 means that the pixels have a linear correlation while -1 denotes a linear anti-correlation. Certainly, there is a better correlation for the sausage than the kink, but this is not surprising since it is clearly visible from Figure \ref{fig:kink} that signal for the kink mode is weaker than for sausage mode (see Figure \ref{fig:Susage}). However, the kink mode stills shows a good correlation in the regions where amplitude is maximum (see Figure \ref{corr}).

\section{Conclusions}
All the methods used to identify coherent oscillations across sunspots and pores have their particular strengths and weaknesses. We have demonstrated here that a more considered and multi-faceted approach can be more robust in pinpointing modes that are actually physical. For example, the previous analysis by J17 required fine tuning the Fourier filters in the temporal and spatial domain to reveal the umbral kink mode confirmed by our POD and DMD analysis. In contrast, POD requires no such filtering, and indeed, such filtering would completely skew the results. The inherent problem with POD is identifying which modes are physical since this method produces as many modes as there are time snapshots. This is where further analysis is required as demonstrated in this study and previously by Higham et al. \cite{higham2018implications}. By calculating the PSD of each POD mode the dominant frequency (or frequencies) of each mode can be identified and these can be paired with the unique frequencies associated each DMD mode allowing for comparison between the spatial structure of the modes produced by both methods. If there is agreement between the spatial structure of both the POD and DMD modes (up to some specified accuracy), then this provides compelling evidence that the mode is indeed physical. To our knowledge, this is the first time the combined approach of using POD and DMD has been used on sunspot data to identify more than one MHD wave mode. We, therefore, suggest that in combination, POD and DMD could prove to be indispensable tools for decomposing the many possible MHD wave modes that could be excited in sunspots and pores, especially with the advent of high resolution observations provided by present and near future ground- and space-based observatories (e.g. Dunn Solar Telescope (DST), Swedish Solar Telescope (SST), The Daniel K. Inouye Solar Telescope (DKIST), Solar-C Space Mission, etc).

\enlargethispage{20pt}

\dataccess{The data used in this paper are from an observing campaign using the Rapid Oscillations in the Solar Atmosphere (ROSA) instrument based at the Dunn Solar Telescope, USA, during December 2011. The Dunn Solar Telescope at Sacramento Peak/NM was operated by the National Solar Observatory (NSO). NSO is operated by the Association of Universities for Research in Astronomy (AURA), Inc., under cooperative agreement with the National Science Foundation (NSF). The NSO historical data archive and its public directory can be found here \href{https://www.nso.edu/data/historical-archive/}{https://www.nso.edu/data/historical-archive/}. ROSA is a six-camera high-cadence solar imaging instrument developed by Queen's University Belfast. Although Queen’s University currently do not have the facilities to host such large data sets publicly for download, the ROSA data archive and ROSA reconstructed data are documented and can be requested and transferred (e.g. FTP or on a physical disk drive) from here \href{https://star.pst.qub.ac.uk/wiki/doku.php/public/research_areas/solar_physics/rosa_archive}{https://star.pst.qub.ac.uk/wiki/doku.php/public/research-areas/solar-physics/rosa-archive}.} 

\aucontribute{GV and VF initiated the overall research in MHD mode identification. ABA and WB carried out the POD and DMD analysis. VF, IB, GV and ABA provided the theoretical background and physical interpretation of obtained results. DJ and MS provided data sets and participated in data interpretation. JH provided his expertise in the methodology of mode decomposition. All the Authors participated in discussing the results and editing the draft.  
}

\competing{The authors declare that they have no competing interests.}

\funding{VF and GV thank to The Royal Society, International Exchanges Scheme, collaboration with Chile (IE170301) and Brazil (IES$\backslash$R1$\backslash$191114), and Science and Technology Facilities Council (STFC) grant ST/M000826/1. This research is also partially funded by the European Union’s Horizon 2020 research and innovation program under grant agreement No. 824135 (SOLARNET). DBJ is grateful to Invest NI and Randox Laboratories Ltd. for the award of a Research \& Development Grant (059RDEN-1).}

\ack{ABA acknowledges the support by Majmaah University (Saudi Arabia) to carry out his PhD studies. VF and GV are thankful for support  provided by The Royal Society and Science and Technology Facilities Council (STFC) and the European Union’s Horizon 2020 (SOLARNET). The authors wish to acknowledge scientific discussions with the Waves in the Lower Solar Atmosphere (WaLSA; www.WaLSA.team) team, which is supported by the Research Council of Norway (project number 262622), and The Royal Society through the award of funding to host the Theo Murphy Discussion Meeting “High resolution wave dynamics in the lower solar atmosphere” (grant Hooke18b/SCTM).}

\bibliographystyle{RS}
\bibliography{RSTA_A}.

\end{document}